\documentclass[fleqn,usenatbib]{mnras}

\usepackage{newtxtext,newtxmath}
\usepackage[T1]{fontenc}

\usepackage{graphicx}
\usepackage{amsmath}
\usepackage{hyperref}
\usepackage{bm}
\usepackage{xcolor}
\usepackage{tikz}

\usepackage{fontawesome}

\title[Constraining $\nu \Lambda$CDM with density-split clustering]{Constraining $\nu \Lambda$CDM with density-split clustering}

\author[E. Paillas et al.]{\parbox{\textwidth}{
Enrique Paillas$^{1, 2}$\thanks{E-mail: enrique.paillas@uwaterloo.ca},
Carolina Cuesta-Lazaro$^{3, 4}$,
Pauline Zarrouk$^{5}$,
Yan-Chuan Cai$^6$,
Will J. Percival$^{1,2,7}$,
Seshadri Nadathur$^8$,
Mathilde Pinon$^9$,
Arnaud de Mattia$^9$,
and Florian Beutler$^6$
}
\vspace*{4pt} \\
\scriptsize $^{1}$ Waterloo Centre for Astrophysics, University of Waterloo, Waterloo, ON N2L 3G1, Canada \\
\scriptsize $^{2}$ Department of Physics and Astronomy, University of Waterloo, 
Waterloo, ON N2L 3G1, Canada \\
\scriptsize $^{3}$ Institute for Computational Cosmology, Department of Physics, Durham University, South Road, Durham DH1 3LE, UK \\
\scriptsize $^{4}$Institute for Data Science, Durham University, South Road, Durham DH1 3LE, UK\\
\scriptsize $^{5}$Sorbonne Universit\'e, Universit\'e Paris Diderot, Sorbonne Paris Cit\'e, CNRS,
Laboratoire de Physique Nucléaire et de Hautes Energies (LPNHE), 4 place Jussieu, F-75252, Paris
Cedex 5, France \\
\scriptsize $^{6}$Institute for Astronomy, University of Edinburgh, Blackford Hill, Edinburgh, EH9 3HJ, UK\\
\scriptsize $^{7}$Perimeter Institute for Theoretical Physics, 31 Caroline St North, Waterloo, ON N2L 2Y5, Canada \\
\scriptsize $^{8}$Institute of Cosmology and Gravitation, University of Portsmouth, Burnaby Road, Portsmouth, PO1 3FX, UK \\
\scriptsize $^{9}$IRFU, CEA, Universit\'e Paris-Saclay, F-91191 Gif-sur-Yvette, France
\vspace*{-2pt}
}

\date{Accepted XXX. Received YYY; in original form ZZZ}
\pubyear{2022}

\begin{document}
\label{firstpage}
\pagerange{\pageref{firstpage}--\pageref{lastpage}}
\maketitle

\begin{abstract}
The dependence of galaxy clustering on local density provides an effective method for extracting non-Gaussian information from galaxy surveys. The two-point correlation function (2PCF) provides a complete statistical description of a Gaussian density field. However, the late-time density field becomes non-Gaussian due to non-linear gravitational evolution and higher-order summary statistics are required to capture all of its cosmological information. Using a Fisher formalism based on halo catalogues from the Quijote simulations, we explore the possibility of retrieving this information using the density-split clustering (DS) method, which combines clustering statistics from regions of different environmental density. We show that DS provides more precise constraints on the parameters of the $\nu \Lambda$CDM model compared to the 2PCF, and we provide suggestions for where the extra information may come from. DS improves the constraints on the sum of neutrino masses by a factor of $7$ and by factors of 4, 3, 3, 6, and 5 for $\Omega_{\rm m}$, $\Omega_{\rm b}$, $h$, $n_s$, and $\sigma_8$, respectively. We compare DS statistics when the local density environment is estimated from the real or redshift-space positions of haloes. The inclusion of DS autocorrelation functions, in addition to the cross-correlation functions between DS environments and haloes, recovers most of the information that is lost when using the redshift-space halo positions to estimate the environment. We discuss the possibility of constructing simulation-based methods to model DS clustering statistics in different scenarios. \href{https://github.com/epaillas/densitysplit-fisher}{\faGithub}
\end{abstract}


\section{Introduction}

In our standard cosmological picture, the $\Lambda$ cold dark matter ($\Lambda$CDM) model, the present large-scale distribution of galaxies evolved from small-scale density perturbations in the early Universe. These perturbations are thought to have originated from quantum fluctuations during a period of inflation, freezing out as a nearly Gaussian random field \citep{Guth1982, Hawking1982}; for a review of primordial non-Gaussianity studies and their implications, see \citet{Desjacques2010}. As such, the statistical properties of the initial density field can be fully characterised by the power spectrum $P(k)$, or, in configuration space, its inverse Fourier transform, the two-point correlation function (2PCF) $\xi(r)$. As the distribution of density fluctuations evolves through gravitational collapse, it becomes non-Gaussian: although overdensities can grow freely, underdensities are always bounded from below, as the density contrast in regions devoid of matter can never go below $\delta = -1$. As a consequence, the density field develops significant skewness and kurtosis, departing from Gaussianity \citep{Einasto2020}. This distribution cannot be completely characterised by the 2PCF anymore, and higher-order correlation functions are needed to describe the density field. Departures from Gaussianity rely on gravity being able to move matter out of its primordial position, so the effect is expected to be less relevant at scales that are much larger than the typical scale of these motions.

Finding summary statistics complementary or supplementary to the 2PCF is now an active area of research in cosmology. Examples include the three-point correlation function \citep{Slepian2017} or bispectrum \citep{Philcox2021a}, the four-point correlation function \citep{Philcox2021b} or trispectrum \citep{Gualdi2021}, counts in cell statistics \citep{Szapudi2004, Klypin2018, Jamieson2020, Uhlemann2020}, non-linear transformations of the density field \citep{Neyrinck2009, Neyrinck2011b, Wang2011, Wang2022:1912.03392v3}, the separate universe approach \citep{Chiang2015}, the marked power spectrum \citep{Massara2018, Massara2022}, the wavelet scattering transform \citep{Valogiannis2021}, void statistics \citep{Hawken2020, Nadathur2020, Correa2020, Woodfinden2022}, density-split gravitational lensing \citep{Gruen2018, Friedrich2018}, and other related statistics. Given that modelling how these statistics change with the cosmological parameters analytically can be challenging and inaccurate on non-linear scales, most studies rely on N-body simulations with varying cosmologies to measure the information content of the statistics in the non-linear regime, such as the Quijote suite of simulations \citep{Villaescusa2020}. For example, \cite{Hahn2020} found that the non-linear redshift-space bispectrum (in particular its monopole) can break degeneracies between cosmological parameters that lead to five times tighter constraints on the sum of neutrino masses, compared to the power spectrum.

Another useful way to retrieve information that leaks to higher orders is by studying galaxy clustering as a function of environmental density \citep{Abbas2007, Tinker2007, Paillas2021, Bayer2021, Bonnaire2022}. Splitting the galaxy field into different density bins naturally captures the non-Gaussian nature of the PDF, and the combination of clustering statistics from different environments can help break parameter degeneracies and improve cosmological constraints \citep{Paillas2021}. As density-split (DS) clustering includes the contribution from underdense regions of the cosmic web, it also shares many of the advantages seen in studies of void statistics. In particular, cosmic voids contain densities of neutrinos higher than those of baryons and dark matter \citep{Massara_2015}. For this reason, void observables are more sensitive to the sum of neutrino masses than two-point statistics \citep{Massara_2015, Kreisch2019}. Here, we show how DS can also access this information and obtain very precise constraints on the sum of neutrino masses.

In this work, we perform a Fisher analysis to quantify the precision with which DS can constrain the value of cosmological parameters in a $\nu \Lambda$CDM model. We study how different definitions of environmental density can affect the constraints of DS and compare them with the results of the standard 2PCF. In particular, we compare the information content of DS when the environments are defined in either real or redshift space. In previous studies \citep{Paillas2021}, several limiting assumptions had to be made to model the clustering of DS multipoles analytically. \cite{Paillas2021} assumed a fixed cosmological template and focused on constraints on the growth rate of structure from redshift-space distortions. Although this highlighted the great potential of DS clustering at extracting non-Gaussian information from galaxy surveys, it did not fully account for the cosmological dependence of the DS correlation functions. To overcome this issue and estimate the full information content of DS, we use the Quijote suite of N-body simulations \citep{Villaescusa2020}, which allows us to explore the cosmological dependence of the full shape of the DS correlation functions. In addition to the cross-correlation functions between DS environments and the tracer field used in \cite{Paillas2021}, we introduce the autocorrelation functions of DS environments, and show that they constitute a valuable source of cosmological information.

The manuscript is organised as follows. In Sect.~\ref{sec:simulations} we describe the simulations used in this work. In Sect.~\ref{sec:ds_algorithm} we describe the density-split clustering algorithm. In Sect.~\ref{sec:fisher_formalism} we outline the main ideas behind the Fisher formalism. We present our main results in Sect.~\ref{sec:main_results}, including an analysis of the information content of density-split clustering in different setups and a comparison against the standard 2PCF. We summarise and present our main conclusions in Sect.~\ref{sec:discussion_conclusions}. We also include an Appendix, where we present various tests that are pertinent for a more in-depth analysis of the results shown in the paper.

\section{The Quijote Simulations} \label{sec:simulations}

The Quijote project \citep{Villaescusa2020} consists of a suite of N-body simulations that were constructed to quantify the information content on cosmological observables. The simulations span a wide range of values around their fiducial cosmology, which is set to a matter density parameter of $\Omega_{\rm m} = 0.3175$, a baryon density of $\Omega_{\rm b} = 0.049$, a dimensionless Hubble constant of $h = 0.6711$, a spectral index of $n_s = 0.9624$, an amplitude of density fluctuations of $\sigma_8 = 0.834$, a neutrino mass of $M_\nu = 0.0\, \rm{eV}$, and a dark energy equation of state of $w = -1$. The fiducial cosmological parameters are in good agreement with the latest Planck constraints \citep{Planck2020}. There are $15,000$ realisations of the fiducial cosmology that can be used to calculate covariance matrices, as well as $500$ realisations of paired simulations where only one cosmological parameter is changed at a time, which can be used to estimate derivatives numerically.

While the initial conditions for most simulations were generated using second-order Lagrangian perturbation theory \citep[2LPT, ][]{Jenkins2010}, the simulations with non-zero neutrino mass were initialised using the Zel'dovich approximation \citep[ZA, ][]{Zeldovich1970}. As we will show later, for a consistent estimation of derivatives with respect to $M_\nu$, we also include simulations of the fiducial cosmology initialised with the ZA (see Sect.~\ref{sec:fisher_formalism} for more details). The specifications of these simulations are listed in Table \ref{tab:simulations}.

Is is worth noting that Quijote provides single- and double-step simulations for calculating derivatives with respect to the baryon density: For $\Omega_{\rm b}^{+}$ and $\Omega_{\rm b}^{-}$, the step is $d\Omega_{\rm b}/\Omega_{\rm b} \sim 2\%$, which produces too small of a difference in our data vectors, making the estimation of the derivatives too noisy and unreliable. For $\Omega_{\rm b}^{++}$ and $\Omega_{\rm b}^{--}$, the step is $d\Omega_{\rm b}/\Omega_{\rm b} \sim 4\%$, which leads to a cleaner estimation of the derivatives in our case, so we use those simulations in this work. For all other cosmological parameters (except $M_\nu$, which is a special case as noted in the paper), only single-step simulations are provided by Quijote, but these produce changes in the multipoles that are large enough to robustly estimate the derivatives.

\begin{table*}
    \centering
    \caption{Characteristics of the Quijote simulations suite that are used in this work. Each row corresponds to a set of simulations with a varying cosmological parameter. The simulations are set to span a grid of cosmologies ready to numerically estimate derivatives with respect to cosmological parameters.}
    \begin{tabular}{l c c c c c c c r}
        \hline
        \hline
        Name & ${\rm \Omega_m}$ & ${\rm \Omega_b}$ & $h$ & $n_s$ & $\sigma_8$ & ${\rm M_{\nu}}$ & realisations & initial conditions\\
        \hline
        Fiducial & 0.3175 & 0.049 & 0.6711 & 0.9624 & 0.834 & 0.0 & 15000 & 2LPT\\
        Fiducial\_ZA & 0.3175 & 0.049 & 0.6711 & 0.9624 & 0.834 & 0.0 & 500 & Zel'dovich approx.\\

        ${\rm \Omega_m^+}$ & 0.3275 & 0.049 & 0.6711 & 0.9624 & 0.834 & 0.0 & 500 & 2LPT\\

        ${\rm \Omega_m^-}$ & 0.3075 & 0.049 & 0.6711 & 0.9624 & 0.834 & 0.0 & 500 & 2LPT\\

        ${\rm \Omega_b^{++}}$ & 0.3175 & 0.051 & 0.6711 & 0.9624 & 0.834 & 0.0 & 500 & 2LPT\\

        $\rm \Omega_b^{--}$ & 0.3175 & 0.047 & 0.6711 & 0.9624 & 0.834 & 0.0 & 500 & 2LPT\\

        $h^+$ & 0.3175 & 0.049 & 0.6911 & 0.9624 & 0.834 & 0.0 & 500 & 2LPT\\

        $h^-$ & 0.3175 & 0.049 & 0.6511 & 0.9624 & 0.834 & 0.0 & 500 & 2LPT\\

        $n_s^+$ & 0.3175 & 0.049 & 0.6711 & 0.9824 & 0.834 & 0.0 & 500 & 2LPT\\

        $n_s^-$ & 0.3175 & 0.049 & 0.6711 & 0.9424 & 0.834 & 0.0 & 500 & 2LPT\\

        $\sigma_8^+$ & 0.3175 & 0.049 & 0.6711 & 0.9624 & 0.849 & 0.0 & 500 & 2LPT\\

        $\sigma_8^-$ & 0.3175 & 0.049 & 0.6711 & 0.9624 & 0.819 & 0.0 & 500 & 2LPT\\

        $M_{\nu}^{+}$ & 0.3175 & 0.049 & 0.6711 & 0.9624 & 0.834 & 0.1 & 500 & Zel'dovich approx.\\
        $M_{\nu}^{++}$ & 0.3175 & 0.049 & 0.6711 & 0.9624 & 0.834 & 0.2 & 500 & Zel'dovich approx.\\
        $M_{\nu}^{+++}$ & 0.3175 & 0.049 & 0.6711 & 0.9624 & 0.834 & 0.4 & 500 & Zel'dovich approx.\\
        \hline
        \hline
    \end{tabular}
    \label{tab:simulations}
\end{table*}

Dark matter halo catalogues in each simulation are generated using a Friends-of-Friends algorithm \citep{1985ApJ...292..371D}. The algorithm works by defining a linking length, which is the maximum distance allowed between particles for them to be considered \textit{friends}. For each particle, the algorithm looks for all other particles within this linking length and groups them together. If two particles are friends with the same particle, they are considered friends with each other and are grouped into the same halo. The process is repeated for all particles until all groups have been identified. In our case, we use a linking-length parameter $b = 0.2$. We select haloes at redshift $z = 0.0$ imposing a minimum halo mass cut of $M_{\rm min} = 3.2 \times 10^{13}\,{h^{-1}\rm M_{\odot}}$, which corresponds to a number density of $n = 1.55 \times 10^{-4}\,(h/{\rm Mpc})^3$. Future surveys, such as DESI \citep{desi}, will be able to sample galaxies living in haloes of much lower masses. Therefore, the constraints shown in this paper do not serve as a forecast for future surveys, but rather serve as a comparison between two-point statistics and DS. 

Adopting a fixed mass cut can modify the bias of the halo samples with respect to the underlying matter distribution, which in turn affects the measured clustering statistics. To disentangle this effect from those coming from variations in cosmological parameters, we also build halo catalogues where we impose mass cuts of $3.1 \times 10^{13}\,{h^{-1}\rm M_{\odot}}$ and ${3.3} \times 10^{13}\,{h^{-1}\rm M_{\odot}}$, so that we can compute derivatives of the data vectors with respect to this mass cut and marginalise over this dependence.

We construct redshift-space halo catalogues by shifting the positions of haloes based on their peculiar velocities along the line of sight (LOS), which is taken to be along one of the axes of the simulation boxes. In most cases, when showing results based on correlation function multipoles, we average the results over 3 different LOS, corresponding to the $x$, $y$, and $z$ axes of the simulations. These three different projections are not fully independent from each other, so when estimating covariance matrices, we only use the projection along the $z$ axis. This results in 1500 realisations of the alternative cosmologies to calculate numerical derivatives. We use 7000 realisations of the fiducial cosmology to estimate covariance matrices.

\section{Density-split clustering} \label{sec:ds_algorithm}

\begin{figure*}
    \centering
    \includegraphics[width=\textwidth]{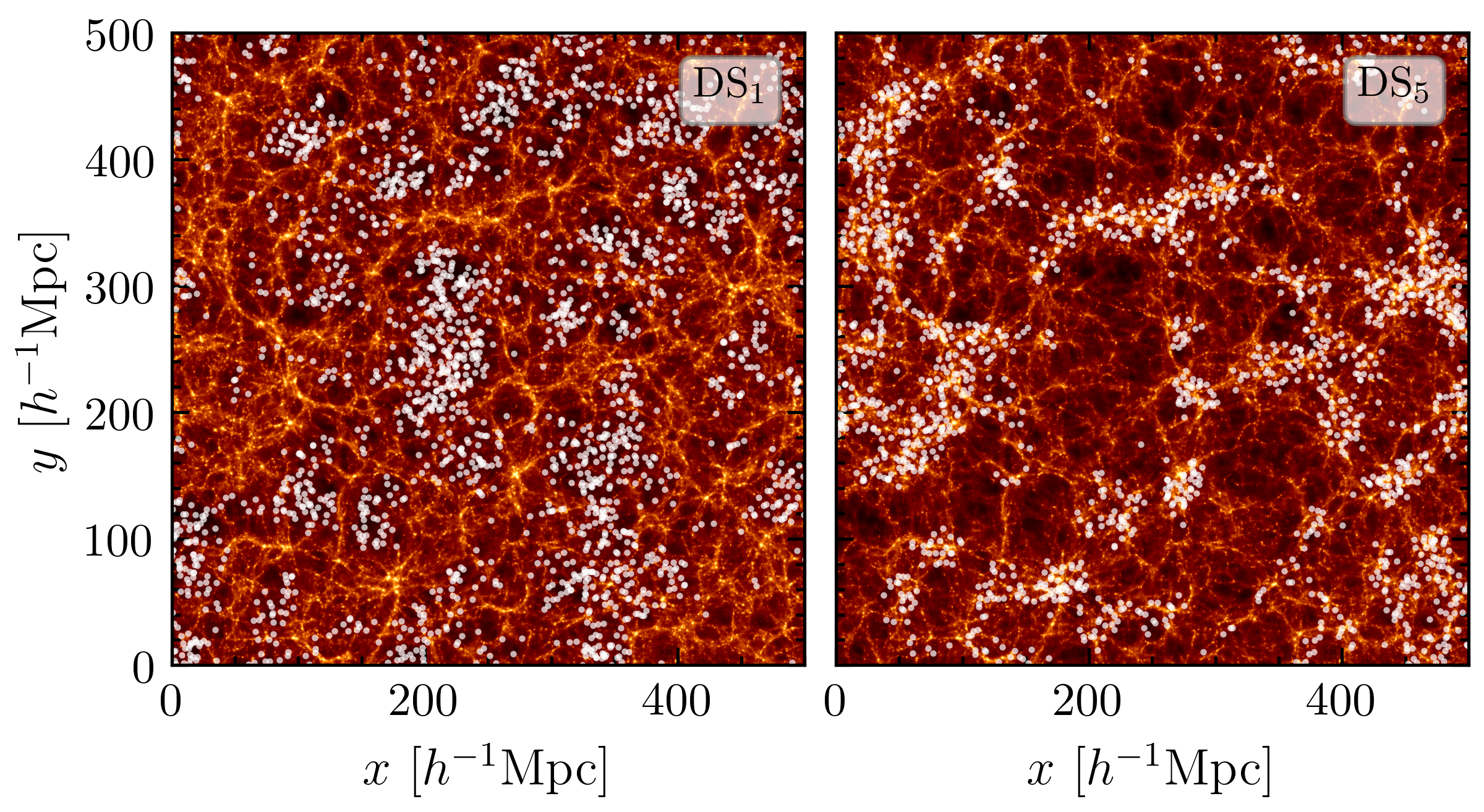}
    \caption{The positions of the ${\rm DS_1}$ and ${\rm DS_5}$ density-split quintiles (white circles) in a region of $500 \times 500 \times 50\, (h^{-1}{\rm Mpc})^3$ from one of fiducial Quijote simulations at $z = 0$. The colourmap shows the projected dark matter density. ${\rm DS_1}$ centres populate the most underdense environments of the cosmic web, whereas ${\rm DS_5}$ centres cluster on high density environments. \href{https://github.com/epaillas/densitysplit-fisher/blob/master/Figure1.py}{\faFileCodeO}}
    \label{fig:projected_density}
\end{figure*}

The density-split clustering method \citep{Paillas2021} consists of splitting a collection of random points according to the local galaxy or halo\footnote{While the algorithm was originally designed to run on galaxies, it can also be applied to catalogues of dark matter haloes or particles.} density contrast at their locations and then extracting cosmological information from the clustering statistics that characterise each environment. We apply the DS algorithm to the dark matter halo catalogues of Quijote simulations using our publicly available code.\footnote{\url{https://github.com/epaillas/densitysplit}} The pipeline can be summarised as follows:

\begin{enumerate}
    \item Generate a set of $N_{\rm random}$ random points that cover the sample volume and measure the integrated halo density contrast $\Delta(R_s)$ in spheres of radius $R_s$ around each random point.
    
    \item Classify the random points into five density bins, or \textit{quintiles}, based on the density contrasts measured from the previous step.
    By definition, each quintile will have the same number of random points. In Fig.\ref{fig:projected_density} we show the random points that were classified as the least ($\mathrm{DS}_1$) and most dense ($\mathrm{DS}_5$) environments in a slice of the Quijote simulations, overlaid on the projected dark matter density in the slice\footnote{The projected dark matter density has been estimated using the DTFE public software (\url{https://github.com/MariusCautun/DTFE}).}. It can be seen that $\mathrm{DS}_1$ points correspond to regions that would usually be denoted as voids, while $\mathrm{DS}_5$ points correspond to nodes of the cosmic web.
    
    \item Measure the multipole moments of the cross-correlation functions between the points in each quintile and the redshift-space halo field, as well as the autocorrelation function of the points in each quintile. The use of the autocorrelations is an addition that was not previously considered in \cite{Paillas2021}. In what follows, we denote autocorrelations of the $i$-th quintile as $\mathrm{DS}_i^{\mathrm{qq}}$ and cross-correlations between the $i$-th quintile and the redshift-space halo field as $\mathrm{DS}_i^{\mathrm{qh}}$.
    
    \item Use the measured multipoles to estimate constraints on the parameters of the $\nu \Lambda$CDM model through a Fisher analysis.
\end{enumerate}

The multipole moments are defined as
\begin{equation}
    \label{eq:multipoles}
    \xi_\ell (s) = \frac{2 \ell + 1}{2} \int_{-1}^1 {\rm d} \mu \, \xi(s, \mu) P_\ell (\mu),
\end{equation}
where $s$ is the pair separation in redshift space, $\mu$ is the cosine of the angle between the separation vector and the line of sight, $P_\ell (\mu)$ are the Legendre polynomials, $\ell=0,2$ for the monopole and the quadrupole, respectively, and $\xi(s, \mu)$ denotes either the cross-correlations between quintiles and the halo field in redshift space, or autocorrelations of quintiles. In principle, valuable information could also be contained in the hexadecapole moment ($\ell=4$), but its statistical uncertainty for the samples considered in this analysis leads to noisy estimates of the numerical derivatives, so we have excluded it from our calculations.

We have run tests with different choices of $N_{\rm random}$, and we have found that the clustering measurements converge when this number is set to five times the number of haloes in each simulation. Therefore, we set $N_{\rm random} = 5 N_{\rm haloes}$ throughout the rest of this work. We set the default smoothing radius $R_s$ to $20\,{h^{-1}{\rm Mpc}}$, which is well above the mean halo separation in the simulations, but still sufficiently small to capture non-Gaussianities in the density PDF. 

The estimation of the halo density around random points in step (i) can be carried out in real or redshift space. \cite{Paillas2021} showed that, from a theoretical point of view, it is easier to model redshift-space multipoles when the density quintiles are defined in real space. However, this can be difficult to apply in real observations, where we only have direct access to the redshift-space galaxy field. A similar problem is found in void-galaxy cross-correlation studies \citep{Nadathur2019c} where reconstruction algorithms \citep{Nadathur2019b} are commonly used to detect voids in real space. However, the reconstruction step also introduces additional complexity when estimating the likelihood of the data given the cosmological parameters, since the reconstructed data depend on some of the parameters being fitted (such as the growth rate of structure, $f$, or the linear galaxy bias). Moreover, reconstruction algorithms are not perfect and might introduce biases in the estimates of real-space quantities that would impact the inference on cosmological parameters. This will be particularly relevant when small scales are included in the analysis, where the signal-to-noise ratio is the largest. Here, we compare both the definitions of the density split and the resulting constraints.

The autocorrelation and cross-correlation functions of each density environment are calculated using \textsc{pycorr}\footnote{\url{https://github.com/cosmodesi/pycorr}}, which is a wrapper around a modified version of CorrFunc \citep{corrfunc}. We use $28$ radial bins within $10\,{h^{-1}{\rm Mpc}} < s < 150\,{h^{-1}{\rm Mpc}}$, and $240$ $\mu$ bins from $-1$ to $1$ for the calculation of redshift-space multipoles. We also measure the multipoles from the halo 2PCF with the same binning scheme for comparison.

Since the distribution of query points that are split by density is random, the sum of the cross-correlation functions of all quintiles vanishes by construction. This means that any of the DS autocorrelation functions can be expressed as a linear combination of the other four quintiles. As a consequence, any combination of four quintiles already contains all the available cosmological information from DS. Therefore, when combining the information from different environments in the likelihood analysis, we do not include the middle quintile, ${\rm DS_3}$. This quintile is the closest to the average density, which makes it less remarkable in terms of its clustering attributes than other quintiles for this particular analysis. However, we have explicitly checked that our cosmological constraints remain largely unaltered if we remove a different quintile from the data vector.

\subsection{The impact of identifying density environments in real or redshift space}

For observational data, we can only access the redshift-space positions of galaxies. However, as in void-galaxy cross-correlation studies, their real space positions can be estimated using reconstruction algorithms \citep{Nadathur2019b}. In this section, we examine the key differences between density splits identified in real ($r$-split), redshift ($z$-split), or reconstructed (\textit{recon}-split) space, and we will later use the Fisher formalism to determine the impact that split identification has on cosmological constraints.

\begin{figure*}
    \centering
    \begin{tabular}{cc}
      \includegraphics[width=0.38\textwidth]{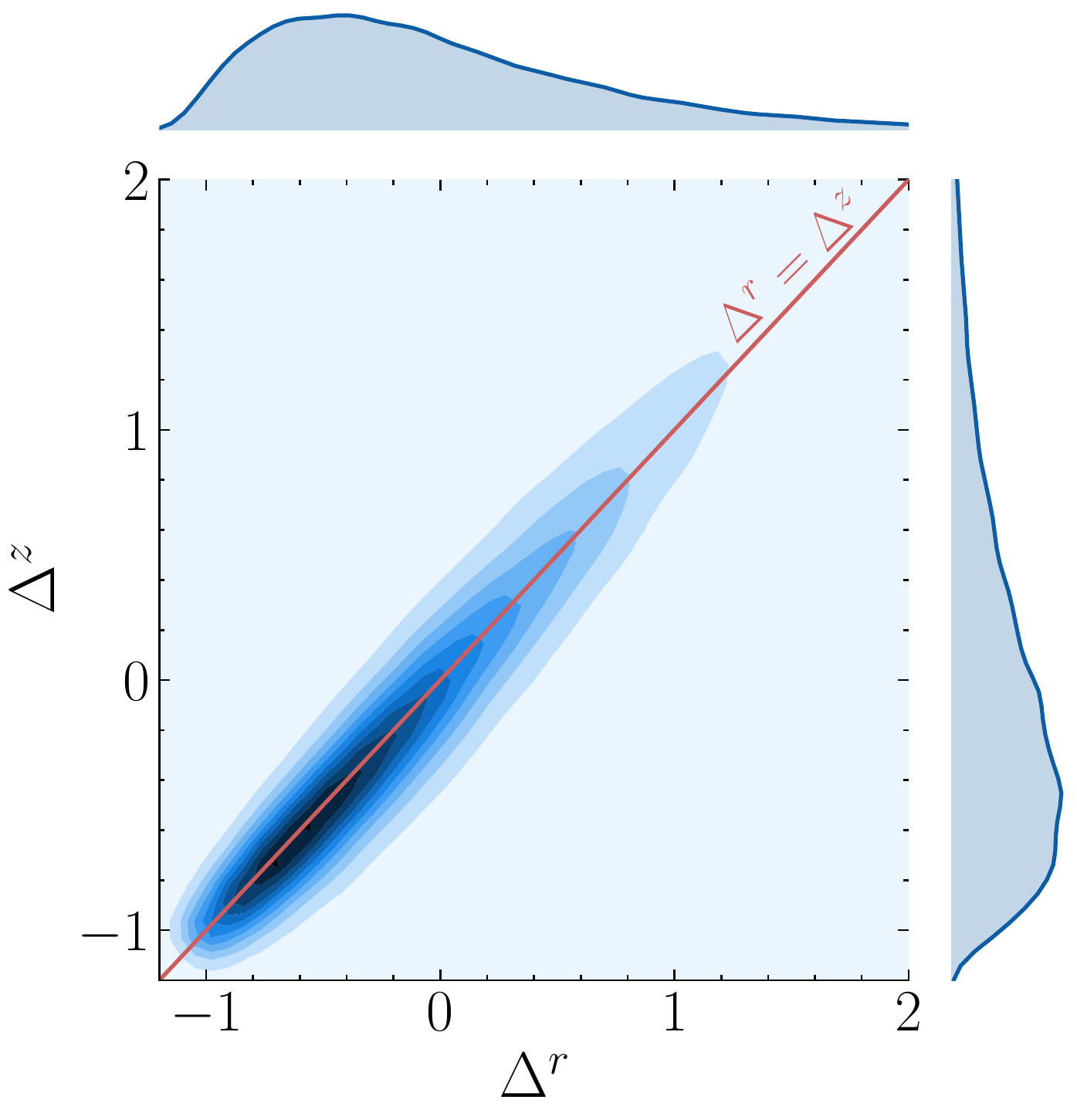}  & \includegraphics[width=0.45\textwidth]{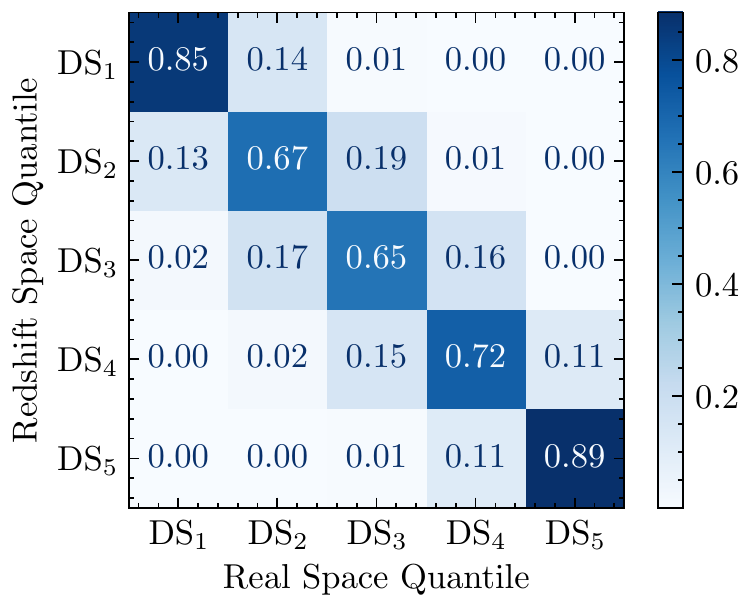}
    \end{tabular}
    \caption{On the left, we show the joint distribution of overdensities, $\Delta$, when identified either in real space $\Delta^\mathrm{r}$, or in redshift space, $\Delta^\mathrm{z}$. In underdense regions, redshift-space densities tend to appear slightly more underdense, whereas overdense regions also appear more overdense. On the right hand side, we show the percent of centres in real space that have been identified as split $i$ but appear as split $j$ in redshift space. \href{https://github.com/epaillas/densitysplit-fisher/blob/master/Figure2a.py}{\faFileCodeO} \href{https://github.com/epaillas/densitysplit-fisher/blob/master/Figure2b.py}{\faFileCodeO}}
    \label{fig:z_vs_r}
\end{figure*}

First, we compare the real and redshift splits using the same set of random centres. This allows us to make a one-to-one comparison of real- and redshift-space environments. In Fig.~\ref{fig:z_vs_r}, we show the joint distribution of overdensities estimated using either the real-space positions of the halos, $\Delta^\mathrm{r}$, or through their redshift-space positions, $\Delta^\mathrm{z}$. The contours are slightly tilted: underdense (overdense) regions appear more underdense (overdense) in redshift space. In underdense regions, outflows of matter will produce deeper density contrasts in redshift space, whereas in overdense regions, coherent infall of matter will tend to produce denser environment estimates.

On the right-hand side of Fig.~\ref{fig:z_vs_r}, we show the percentage of random points that belong to a given quintile in real and redshift space. When the density split is performed in redshift space, a substantial fraction of each quintile consists of misclassified points, which would have been part of a different quintile based on their true (real-space) density. This misclassification mostly shifts points from one quintile to its nearest neighbour(s), and larger shifts are rare.

We will now focus on the effect that this has on the multipoles of autocorrelations and cross-correlations.

\begin{figure*}
    \centering
    \begin{tabular}{cc}
      \includegraphics[width=0.45\textwidth]{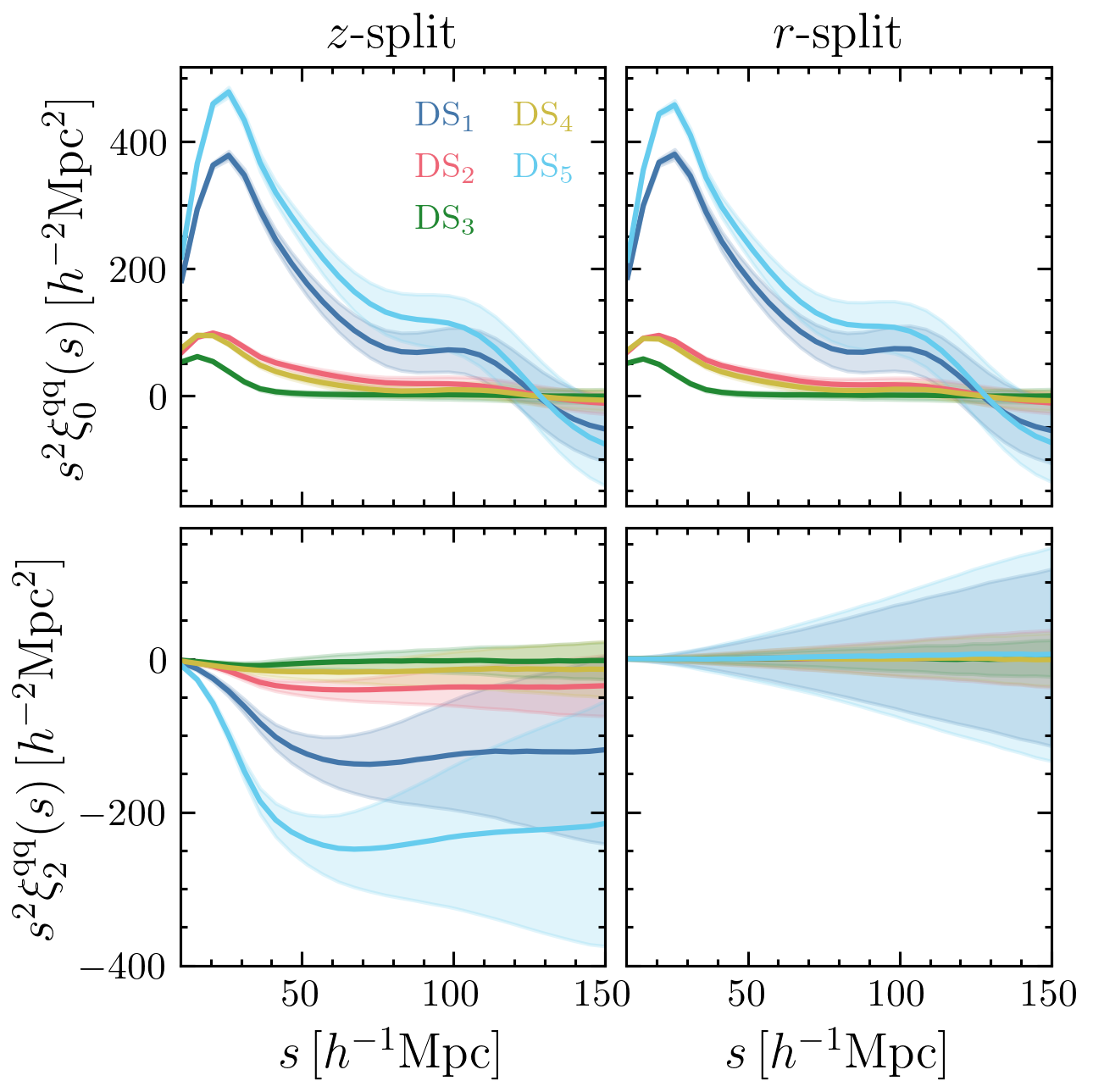}  & \includegraphics[width=0.45\textwidth]{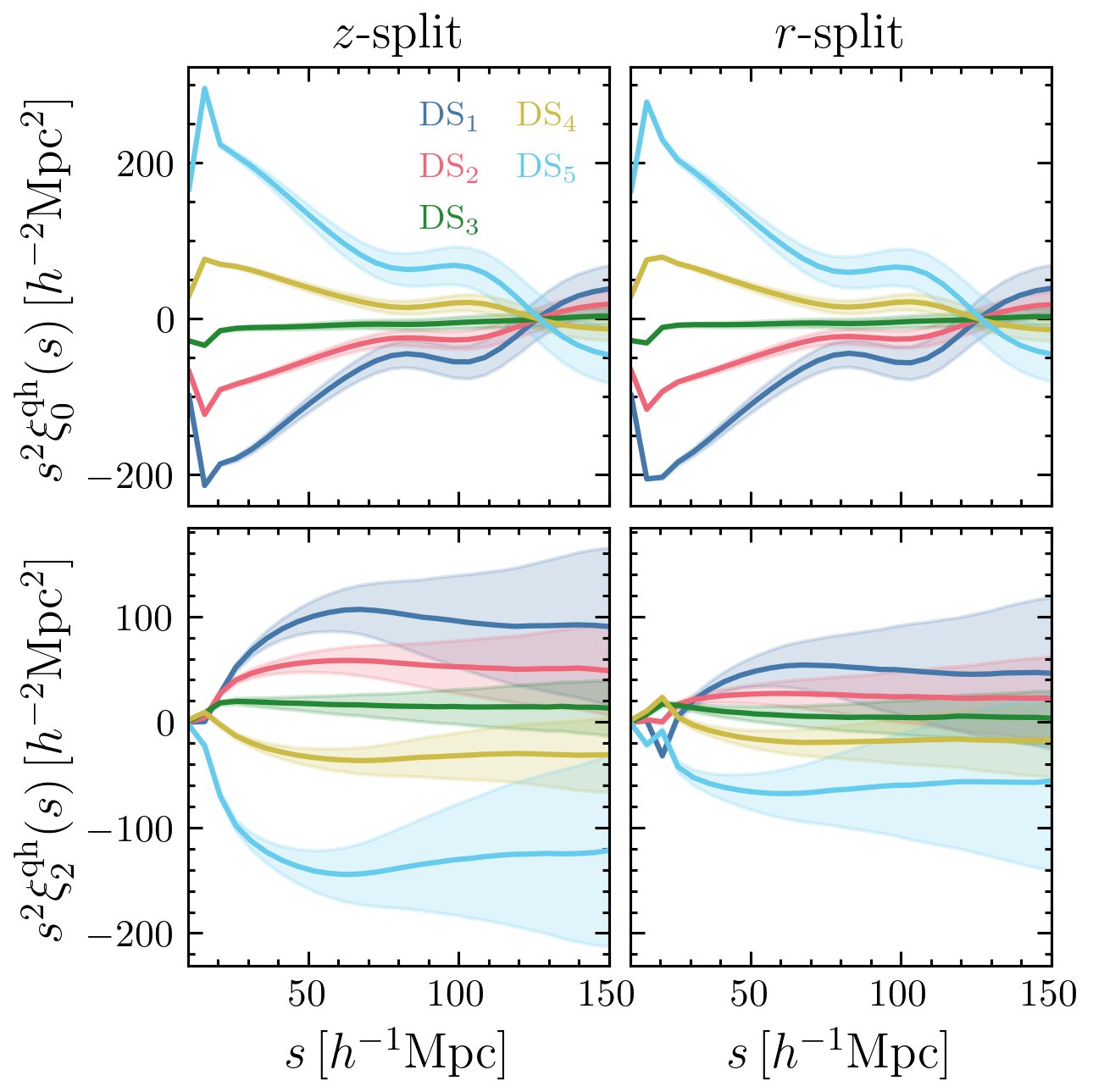}
    \end{tabular}
    \caption{Multipoles of the DS autocorrelation functions (left panel) and DS-halo cross-correlation functions (right panel). The subpanels compare the cases when the quintiles are defined in redshift or real space (left and right sub-panels, respectively). Error bars represent the standard deviation associated to a $(1\,h^{-1}{\rm Gpc})^3$ volume, estimated from multiple mock realizations of the fiducial cosmology. \href{https://github.com/epaillas/densitysplit-fisher/blob/master/Figure3.py}{\faFileCodeO}}
    \label{fig:ds_multipoles}
\end{figure*}

Figure \ref{fig:ds_multipoles} shows the multipoles of DS cross-correlation ($\mathrm{DS}_i^{\mathrm{qh}}$) between points in a quintile and the halos' redshift-space positions, and autocorrelation ($\mathrm{DS}_i^{\mathrm{qq}}$) functions of random points within the same quintile, when the overdensities are estimated from the real-space positions of halos ($r$-split) or from their redshift-space positions ($z$-split).

For the autocorrelations, shown on the left-hand side of Fig.~\ref{fig:ds_multipoles}, the monopole is very similar in both the real-space and redshift-space splits. In both cases, the largest signal is found for the overdense regions $\mathrm{DS}_5$, closely followed by the underdense regions $\mathrm{DS}_1$. We note that even though $\mathrm{DS}_1$, $\mathrm{DS}_2$ and $\mathrm{DS}_3$ are expected to have a negative tracer bias due to their underdense nature, all autocorrelation monopoles are positive since the bias enters as its square in the mapping from matter to tracer autocorrelation functions, i.e., $\xi_{\rm tracer} = b^2 \xi_{\rm matter}$. Both $\mathrm{DS}_1$ and $\mathrm{DS}_5$ show a significant enhancement in clustering on a scale of $\sim 100\,h^{-1}{\rm Mpc}$ corresponding to the acoustic scale set by the baryon acoustic oscillations (BAO). The other quintiles also feature the BAO signal at the same scale, although it is harder to notice because of their smaller amplitudes.

The quadrupole, on the other hand, is completely different for the real-space and redshift-space identification scenarios. It is compatible with zero for splits identified in real space, whereas it is always negative for splits done in estimated redshift-space densities. In the $r$-split scenario, where density splits are performed in real space, there is no preferred direction, and so statistical isotropy dictates a quadrupole signal consistent with zero. When estimating densities in redshift space, peculiar velocities along the line of sight introduce a direction-dependent distortion to the estimated density field, which creates a redshift-space distortion (RSD) anisotropy in the distribution of the DS centres themselves. As discussed earlier, RSD results in a misclassification of some of the random points, which tend to swap to their neighbouring quintile in redshift space. This misclassification occurs in an anisotropic way, which leads to a distorted clustering pattern of the quintile centres. In Appendix~\ref{ap:r_vs_z}, we explicitly show how these misidentifications contribute to the quadrupole by decomposing it into the contributions from the correctly identified and misidentified centres.  Generally, a non-linear transformation of a tracer density field performed in redshift space, such as large-scale structure identification from halo catalogues, will itself have RSD, with an additional velocity bias \citep{Seljak2012, Chuang2017}.

We caution the reader against interpreting the differences between $z$-split and $r$-split identified DS multipoles based on the inferred motion of the random centres. One could define a velocity to be associated with the random centres based on the average velocities of the dark matter particles within the spheres that were used to estimate the environment density, and then use these velocities to map $r$-split multipoles into $z$-split ones. However, this interpretation could not explain the negative quadrupole of negatively biased density split centres such as $\mathrm{DS}_1$. A negative bias implies a positive mean pairwise velocity on large scales, which would produce a positive quadrupole by elongating the two-point correlation function along the line of sight. We have also used the same random seed when generating the random points for performing the $r$-split and $z$-split, so that the random centres by construction have the same positions in real and redshift space, and no motion occurs. One should therefore avoid thinking about the motion of the random centres since in the DS pipeline the random centres are never moved but simply re-classified into different quintiles in the $z$-split scenario. One should instead interpret the anisotropies in the correlation function in terms of how the same random centres are classified in either real or redshift space.

On the right-hand side of Fig.~\ref{fig:ds_multipoles}, we show the multipoles resulting from cross-correlating the random centres in each quintile with the halos' redshift space positions. In the left column, we show the cross-correlation with centres identified in redshift space, whilst on the right we show the same cross-correlation when centres are identified in real space. In both cases, the halo positions are in redshift space. The monopole moment, which appears to be largely unaffected by the density split definition, shows a wide range in amplitudes at small scales, going from the most underdense regions in ${\rm DS}_1$, having density contrasts close to $-1$, to the overdense environments of ${\rm DS}_5$, which correspond to cluster-like environments with density contrasts around 2. These amplitudes also reflect the non-Gaussian nature of the density PDF: ${\rm DS}_1$ regions are always constrained from below, as voids cannot be emptier than empty ($\delta = -1$). However, the densities in ${\rm DS}_5$ can go well beyond 1, breaking the symmetry of the distribution. At large scales, the monopole moments slowly converge towards the mean density. In a Gaussian random field, the splits would be perfectly symmetric; deviations from it are a signature of non-Gaussianity in the density field. Around the scale of $100\,h^{-1}{\rm Mpc}$ we can distinguish the signal coming from BAO for all density quintiles, both for the cross-correlation and autocorrelation functions. 

Regarding the quadrupole moment of the cross-correlations, they show features that can be very different between the two identification scenarios. On large scales, where the two cases behave qualitatively similarly, we see positive amplitudes in ${\rm DS}_1$, ${\rm DS}_2$ and ${\rm DS}_3$, while negative amplitudes are observed in ${\rm DS}_4$ and ${\rm DS}_5$. According to our convention for the redshift-space multipoles [Eq.~(\ref{eq:multipoles})], a negative (positive) quadrupole for overdensities (underdensities) means that the distribution of haloes around these quintiles appears to be flattened along the line of sight. We also observe that the amplitudes of the quadrupoles for ${\rm DS}_1$ and ${\rm DS}_5$ are higher in $z$-split than in $r$-split. This is again a consequence of the misidentification of quintiles and the additional anisotropy that the redshift-space definition of quintiles introduces.

For the redshift-space identification scenario, the quadrupoles maintain their sign across the whole scale range. However, for the real-space identification, we see an abrupt change from positive to negative amplitudes for ${\rm DS_1}$. This transition, which translates to an apparent elongation of the underdensities along the line of sight, has also been observed in the void-galaxy cross-correlation function \citep{Nadathur2020, Woodfinden2022}, and can be driven by the coherent outflow of galaxies from voids \citep[see][for a more in-depth discussion about the physical interpretation of this feature]{Cai2016, Nadathur2019a}.

\subsection{Reconstructing real-space positions}
\label{sec:reconstruction}

\citet{Nadathur2019b} proposed to detect voids after reconstructing the approximate real-space galaxy positions by removing the effects of large-scale velocity flows from the redshift-space positions. The reconstruction algorithm is similar to that used in BAO analyses \citep{Padmanabhan2012, Bautista2018, Chen2022}, but is employed only to remove the RSD, not to remove non-linearities in the density field. This is motivated by the theoretical challenges that arise from modelling the clustering around cosmic voids when these are identified from redshift-space galaxy catalogues. By using a density-field reconstruction algorithm, they were able to move galaxies back to their approximate real-space positions, which can then be used to identify voids.  Here, we use the same method to remove RSD from the redshift-space Quijote halo catalogues and then identify the DS quintiles in the reconstructed catalogues.

Let us place ourselves in a Lagrangian framework, in which the Eulerian position $\vec{x}$ at time $t$ can be described in terms of the initial Lagrangian position $\vec{q}$ and a non-linear displacement field $\vec{\Psi}(\vec{q}, t)$:
\begin{equation}
    \vec{x}(\vec{q}, t) = \vec{q} + \vec{\Psi}(\vec{q}, t)\, .
\end{equation}
The halo overdensity field $\delta_{h}(\vec{x},t)$, can be related to the displacement field by \citep{Nusser1994}
\begin{equation} \label{eq:reconstruction}
    \nabla \cdot \vec{\Psi} + \frac{f}{b} \nabla \cdot (\vec{\Psi} \cdot \hat{r})\hat{r} = -\frac{\delta_{h}}{b}\, ,
\end{equation}
where $b$ is the linear bias of the halo sample. The full solution to Eq.~(\ref{eq:reconstruction}) includes contributions to the velocity flow coming from galaxy peculiar velocities at the corresponding redshift, as well as additional non-linear evolution that can be traced back to earlier epochs. In BAO analyses \citep[e.g.][]{Alam2017}, in an attempt to undo all effects of non-linear clustering to sharpen the BAO feature to the best extent possible, galaxy or halo positions are shifted by $-\vec{\Psi}$ using the full displacement field.

In our analysis, we are only concerned with removing the RSD coming from halo peculiar velocities at a certain epoch, so the part of the solution we are interested in is
\begin{equation} \label{eq:reconstruction_rsd_solution}
    \vec{\Psi}_{\rm RSD} = -f (\vec{\Psi} \cdot \hat{r})\hat{r}\, .
\end{equation}
Shifting the redshift-space halo positions by $-\vec{\Psi}_{\rm RSD}$, we obtain a pseudo real-space halo catalogue that can be used to define the DS quintiles.

Several reconstruction implementations have been introduced in the literature. Here, we use the Iterative FFT Particle Reconstruction code implemented in \textsc{pyrecon},\footnote{\url{https://github.com/cosmodesi/pyrecon}} which solves Eq.~(\ref{eq:reconstruction}) by using an iterative Fast Fourier Transform (FFT) procedure \citep{Burden2015}. This is the same algorithm that was applied to reconstruct the galaxy field in the eBOSS cosmological analysis \citep{Bautista2018}, and for reconstruction in void-galaxy cross-correlation measurements \citep{Nadathur2019c,Nadathur2020, Woodfinden2022}. Eq.~(\ref{eq:reconstruction}) shows that reconstruction is sensitive to the ratio of the linear growth rate of structure $f$ and the linear bias parameter $b$. We estimate the value of $f$ from the cosmology of the fiducial Quijote simulation as $f = \Omega_{\rm m}(z) ^ {0.55} = 0.532$. We estimate the linear halo bias taking the square root of the ratio between the halo and the matter power spectrum, which yields a value of $b = 1.7$ at large scales. The FFT procedure operates on the density field on a regular grid, which we set to have a size of $512^3$. The density field $\delta_h$ is smoothed with a Gaussian kernel of width $R_s^{\rm recon}$ to reduce sensitivity to small-scale density modes, for which Eq.~(\ref{eq:reconstruction}) becomes inaccurate. We adopt $R_s^{\rm recon} = 10\,h^{-1}{\rm Mpc}$, in line with \cite{Nadathur2020} for easier comparison.

\begin{figure}
    \centering
    \includegraphics[width=0.47\textwidth]{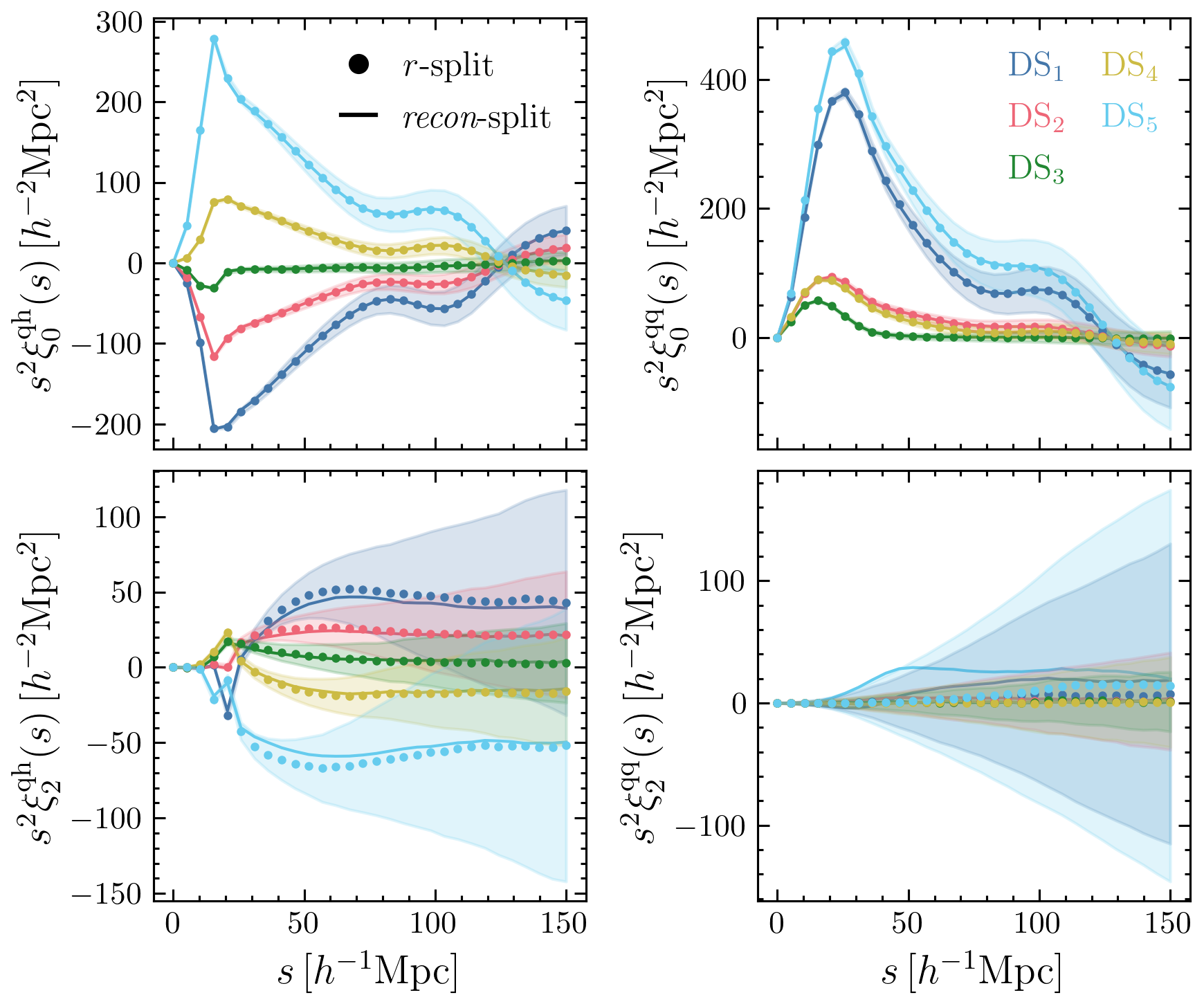}
    \caption{Comparison of multipoles when the densities are identified in either real (dots) or reconstructed halo positions (lines). Left: DS-halo cross-correlation functions. Right: DS autocorrelation functions. Shaded regions represent the standard deviation associated to a $(1\,h^{-1}{\rm Gpc})^3$ volume, estimated from multiple mock realisations of the fiducial cosmology. \href{https://github.com/epaillas/densitysplit-fisher/blob/master/Figure4.py}{\faFileCodeO}}
    \label{fig:reconstructed_multipoles}
\end{figure}

We show the multipoles obtained when splitting the density field using the reconstructed real-space positions of halos (\textit{recon}-split) in Fig.~\ref{fig:reconstructed_multipoles}, where we also compare against the real-space identification scenario ($r$-split). Qualitatively, we find that the \textit{recon}-split multipoles closely follow the key features observed in the $r$-split multipoles: i) the quadrupole of the autocorrelation functions being consistent with zero, ii) the smaller amplitudes of the cross-correlation functions' quadrupole with respect to the $z$-split case, and iii) the transition from a positive to negative quadrupole for the ${\rm DS}_1$ cross-correlation function. Overall, we find that the \textit{recon}-split multipoles lie within 1-$\sigma$ of the $r$-split multipoles for a wide range of scales, although some deviations can be seen in the quadrupole of ${\rm DS}_1$ and ${\rm DS}_5$ around $\sim 50\,h^{-1}{\rm Mpc}$. In the next section, we will assess whether we can recover unbiased constraints for the cosmological parameters using \textit{recon}-split multipoles if we model them as if they were $r$-split measurements.

\section{Fisher formalism} \label{sec:fisher_formalism}

We quantify the information content of the summary statistics using the Fisher formalism \citep{Fisher1935, Tegmark1997, Tegmark1997b}. Given a set of model parameters $\theta$ (in our case, the parameters of the $\nu\Lambda$CDM cosmological framework), we can measure the information on $\theta$ carried by an observed data vector $\bm{d}$ (in our case, the 2PCF or DS multipoles) by calculating the Fisher matrix, defined as
\begin{equation}
    \mathcal{F}_{ij}(\bm{\theta}) = \left \langle \left( \frac{\partial }{\partial \theta_i} \log \mathcal{L}(\bm{d}|\bm{\theta}) \right) \left(\frac{\partial }{\partial \theta_j} \log \mathcal{L}(\bm{d}|\bm{\theta}) \right)  \right \rangle_{\bm{d}}
    \label{eq:definition_fisher}
\end{equation}
where $\mathcal{L}(\bm{d}|\bm{\theta})$ is the likelihood of the data vector given the parameters $\theta$. We note that the expectation is taken over the data, since it is itself a random variable.

The derivative of the likelihood with respect to the parameters is also known as the score function $s(\bm{\theta}) =  \frac{\partial }{\partial \bm{\theta}} \log \mathcal{L}(\bm{d}|\bm{\theta})$, which is zero at the maximum likelihood point. Eq.~(\ref{eq:definition_fisher}) can be interpreted as the variance of the score function, since the expected value of the score function is zero. A random variable that contains high Fisher information implies that the absolute value of the score is often high. Fisher information is used to quantify the effect that small changes in $\bm{\theta}$ have on the likelihood. If small changes in $\theta$ substantially vary the likelihood, then we will be able to set tight constraints on the parameters, and we say that the information content of $\bm{d}$ in $\bm{\theta}$ is large.

When the likelihood can be differentiated twice, it can be shown that the variance of the score is also related to the second derivative, and therefore to the curvature, of the likelihood function
\begin{equation}
    \mathcal{F}_{ij}(\bm{\theta}) = - \left \langle  \frac{\partial ^2 }{\partial \theta_i \partial \theta_j } \log \mathcal{L}(\bm{d}|\bm{\theta}) \right \rangle,
\end{equation}
implying that a more peaked likelihood contains more information on the parameters than a flatter one.

The Cramér–Rao bound states that the inverse of the Fisher information is a lower bound on the variance of any unbiased estimator of $\bm{\theta}$
\begin{equation}
\sigma_{\theta_i} \ge \sqrt{(\mathcal{F}^{-1})_{i, i}}\, .
\label{eq:cramer_rao}
\end{equation}

In particular, if the likelihood follows a multivariate Gaussian distribution, we can compute the expectation value in the calculation of the Fisher matrix analytically, finding
\begin{equation}
\label{eq:fisher_gaussian}
    \mathcal{F}_{ij}(\bm{\theta}) = \frac{1}{2} \mathrm{Tr} \left[ C^{-1} \frac{\partial C}{\partial \theta_i} C^{-1}\frac{\partial C}{\partial \theta_j} + C^{-1}\left( \frac{\partial{\bm d}}{\partial \theta_i} \frac{\partial {\bm d}}{\partial \theta_j}^\top  + \frac{\partial{\bm d}}{\partial \theta_i}^\top \frac{\partial {\bm d}}{\partial \theta_j} \right) \right],
\end{equation}
where $C$ is the covariance matrix associated with the data vector $\bm{d}$. As shown by \citet{refId0}, the first term in Eq.~(\ref{eq:fisher_gaussian}) artificially adds information that was already included in the second term through the derivative of the mean vector. In the rest of the paper, we neglect this term to rather produce a conservative estimate of the information content and compute the Fisher matrix as
\begin{equation}
    \label{eq:fisher}
    \mathcal{F}_{ij}(\bm{\theta}) = \frac{\partial {\bm d}}{\partial \theta_i}C^{-1} {\frac{\partial {\bm d}}{\partial \theta_j}}^\top .
\end{equation}

In Appendix~\ref{ap:gaussianity_likelihood}, we show that the likelihood for DS statistics is indeed very close to a multivariate Gaussian. We note that non-Gaussianities in the likelihood could lead to artificially tight bounds on the cosmological parameters when using the Fisher matrix formalism described by Eq.~(\ref{eq:fisher}) \citep{Park2023}.

For most of the cosmological parameters, the derivatives can be numerically approximated as
\begin{equation} \label{eq:derivatives}
    \frac{\partial \bm{d}}{\partial \theta_i} \simeq \frac{{\bm d}(\theta_i + {\rm d}\theta_i) - {\bm d}(\theta_i - {\rm d}\theta_i)}{2 {\rm d}\theta_i}\,,
\end{equation}
which is a second-order approximation in $\theta_i$. Eq.~(\ref{eq:derivatives}) cannot be used to estimate the derivatives with respect to $M_\nu$, as the neutrino mass cannot be negative. In that case, we instead approximate it as follows:
\begin{equation} \label{eq:derivatives_neutrinos}
    \frac{\partial {\bm d}}{\partial M_\nu} \simeq \frac{\bm{d}(4  \mathrm{d}M_\nu) - 12\bm{d}(2\mathrm{d}M_\nu) + 32\bm{d}(\mathrm{d}M_\nu) - 21\bm{d}(M_\nu = 0)}{12 \mathrm{d}M_\nu}\,,
\end{equation}
which is of second order in $M_\nu$. For a consistent estimation of the derivatives, the $M_\nu = 0$ data vector in Eq.~(\ref{eq:derivatives_neutrinos}) is measured from simulations of the fiducial cosmology with initial conditions generated using the Zel'dovich approximation (see Sect.~\ref{sec:simulations}).

\begin{figure*}
    \centering
    \begin{tabular}{cc}
      \includegraphics[width=0.4\textwidth]{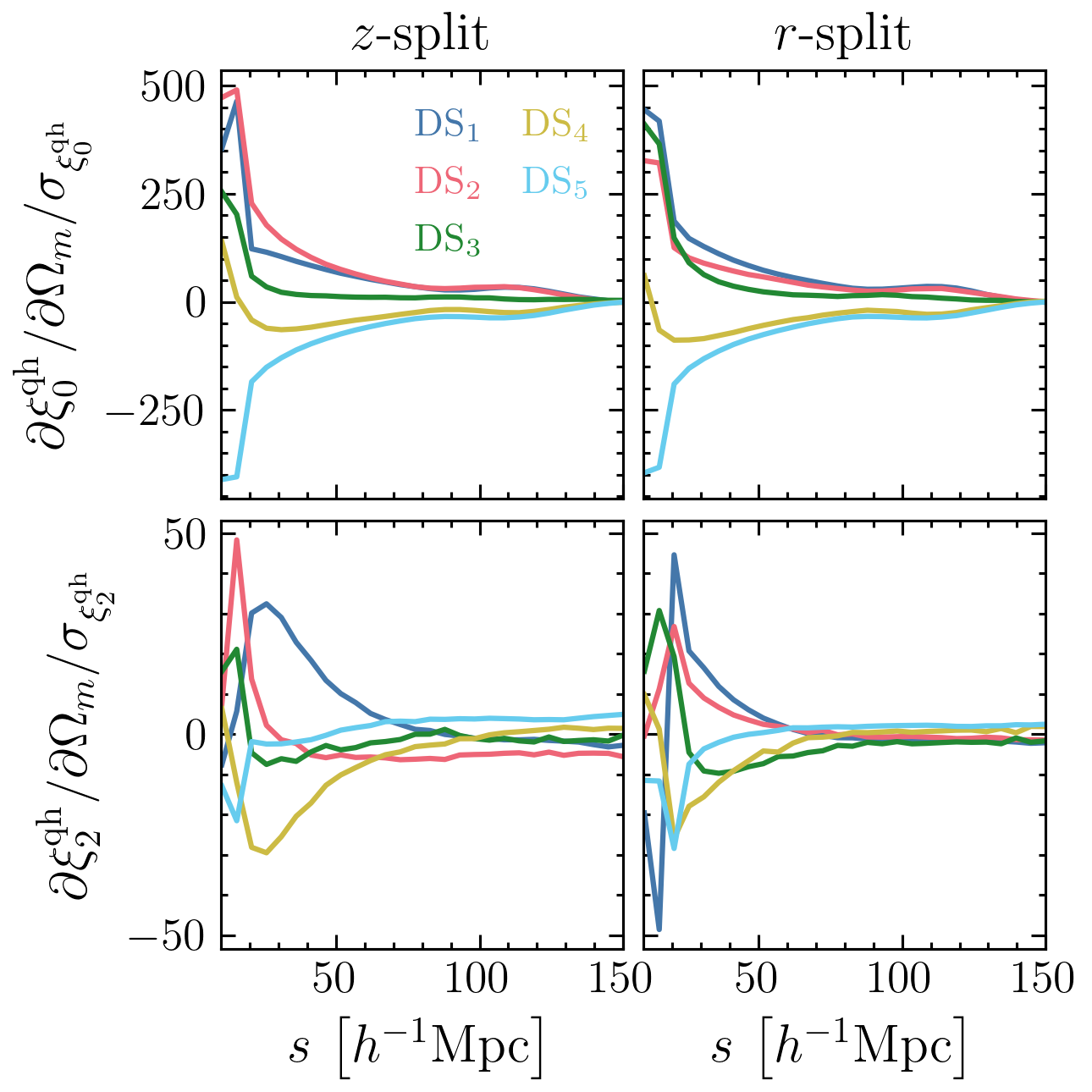}  & \includegraphics[width=0.4\textwidth]{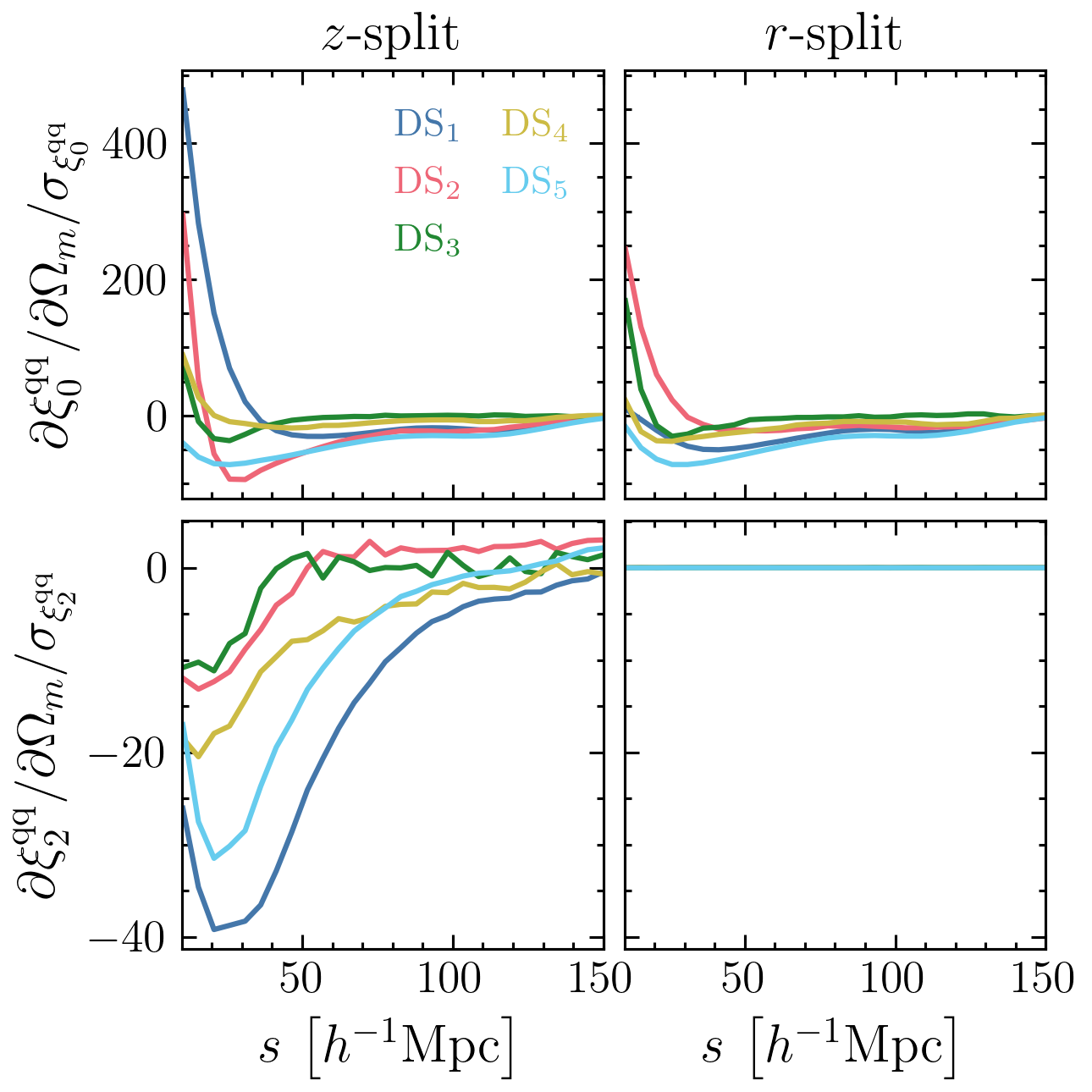}
    \end{tabular}
    \caption{(left) derivatives of the DS-halo cross-correlation multipoles with respect to $\Omega_{\rm m}$, expressed in units of the variance of the multipoles. The upper and lower rows in each panel show derivatives of the monopole and quadrupole moments, respectively, while the left and right columns compare results when the quintiles are defined in redshift or real space. (right) same as the other panel, but showing the DS autocorrelation functions. \href{https://github.com/epaillas/densitysplit-fisher/blob/master/Figure5.py}{\faFileCodeO}}
    \label{fig:derivatives}
\end{figure*}

We calculate derivatives of the redshift-space 2PCF and DS multipoles on each of the 500 realisations of the paired simulations along three different lines of sight (taken to be the $x$, $y$ and $z$ axes of the simulations), which effectively gives us 1500 realisations over which we take the average \citep{Smith2020}. Figure \ref{fig:derivatives} shows an example of these derivatives for the matter density parameter, $\Omega_{\rm m}$. Each quintile shows a distinct sensitivity to $\Omega_{\rm m}$ as a function of scale. The largest contribution comes from small scales, where we expect the density field to deviate the most from a Gaussian distribution. The auto- and cross-correlation functions also show different scale dependencies, which, as we will corroborate later, highlight the importance of combining these two sets of statistics to maximise the cosmological constraining power. We also note that the contribution from the quadrupole of the $r$-split autocorrelations is consistent with zero, which agrees with the discussion presented in the previous section, where we showed that in this scenario the centres of the DS quintiles are distributed isotropically in the simulation volume.

We estimate the covariance matrix from the multiple realisations of the fiducial cosmology as
\begin{align} \label{eq:covariance}
    C = \frac{1}{n_{\rm sim} - 1} \sum_{k=1}^{n_{\rm sim}} \left({\bm{d}_{k}} - \overline{\bm{d}}\right)\left({\bm{d}_{k}} - \overline{\bm{d}}\right) \, ,
\end{align}
where $n_{\rm sim}=7000$ and $\overline{\bm{d}}$ is the mean data vector averaged over all the realisations. In Appendix~\ref{ap:convergence_tests} we show that the inferred errors on the parameters converge when using these numbers of realisations for the calculation of the derivatives and covariance.

In order to obtain the parameter constraints, two matrix inversions need to be performed: the inversion of the covariance matrix in Eq.~(\ref{eq:fisher}), and that of the Fisher matrix in Eq.~(\ref{eq:cramer_rao}). Although the estimator of the covariance matrix [Eq.~(\ref{eq:covariance})] is unbiased, these two inversions lead to biased constraints on the parameters. To account for this, we apply a correction to the covariance matrix 
\begin{equation}
  C' = \frac{n_{\rm sim}-1}{n_{\rm sim} - n_{\rm bins} + n_\theta - 1}C\,,
\end{equation}
where $n_{\theta}$ is the number of parameters and $n_{\rm bins}$ is the number of bins in the data vector. The derivation of this correction factor is presented in Appendix~\ref{ap:covariance_correction}. 

Figure \ref{fig:covariance} shows the correlation matrix associated with this covariance for the $z$-split DS and 2PCF data vectors. For DS, the covariance includes contributions from the monopole and quadrupole moments of the auto and cross-correlation functions for each for the DS quintiles. Since we use 28 radial bins in the range $10 < s < 150 {h^{-1}{\rm Mpc}}$, this results in a $560 \times 560$ matrix. For the 2PCF, we have a $56 \times 56$ matrix resulting from the contributions from the monopole and quadrupole.

\begin{figure*}
    \centering
    \begin{tabular}{cc}
      \includegraphics[width=0.4\textwidth]{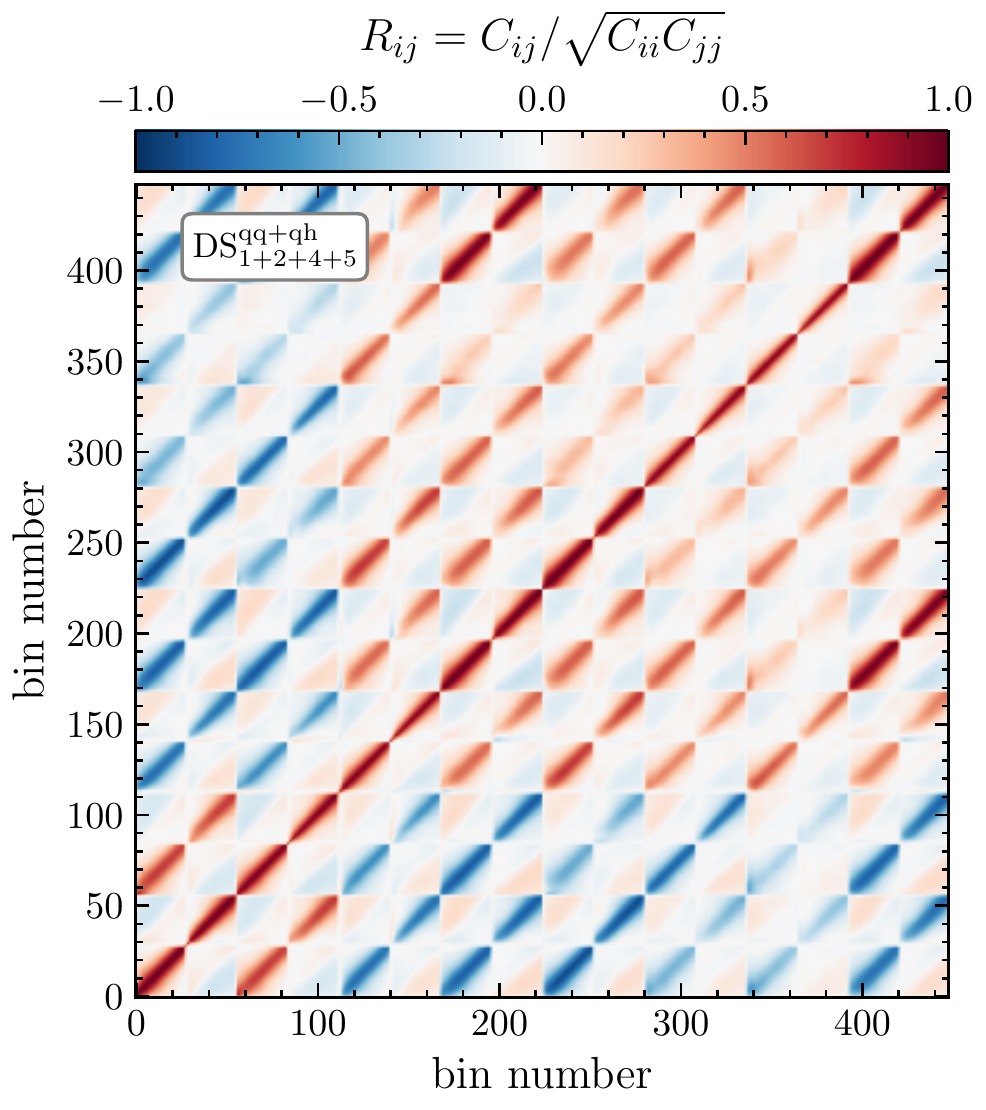}  & \includegraphics[width=0.395\textwidth]{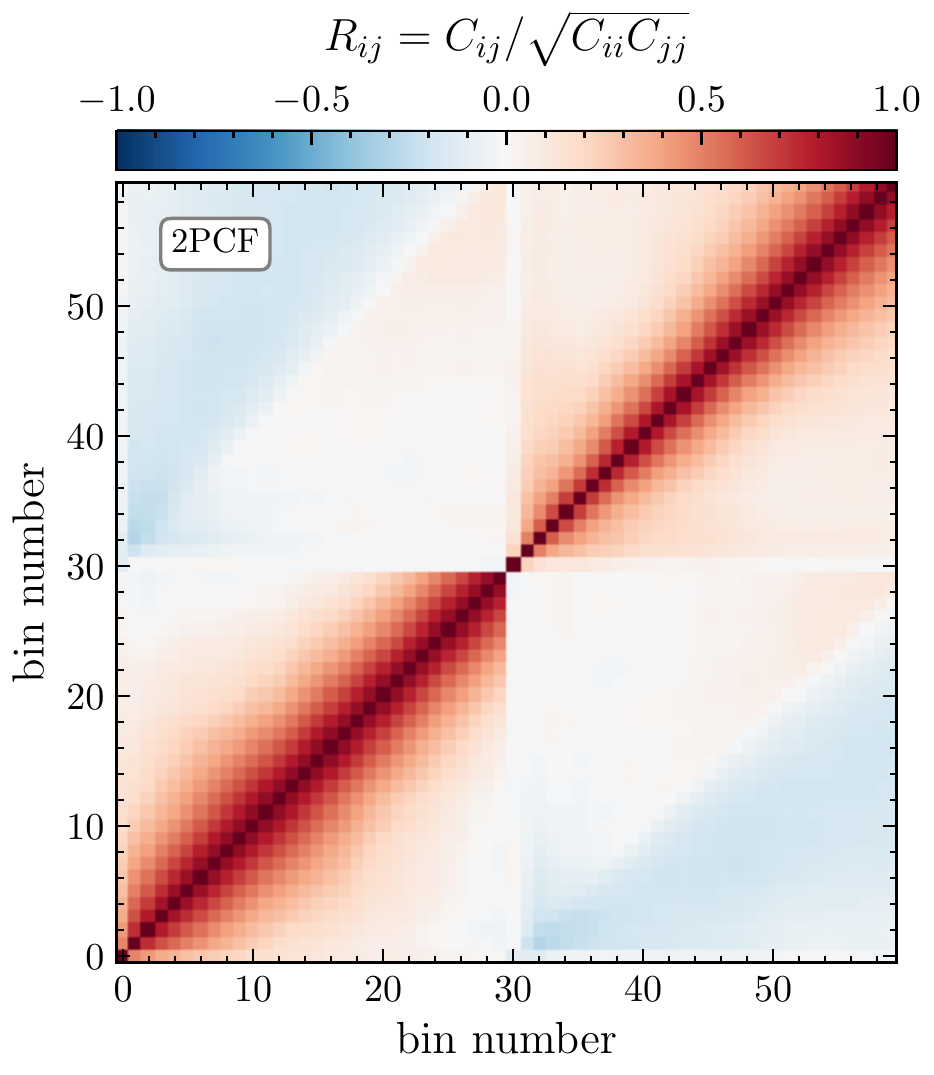}
    \end{tabular}
    \caption{Correlation matrices of the DS and 2PCF data vectors, which include contributions from the monopole and quadrupole moments of the redshift-space correlation functions. \href{https://github.com/epaillas/densitysplit-fisher/blob/master/Figure6.py}{\faFileCodeO}}
    \label{fig:covariance}
\end{figure*}

\section{Information content of density-split clustering} \label{sec:main_results}

\subsection{Identifying environments} \label{subsec:identifying_environments}

The first step of the DS algorithm described in Sect.~\ref{sec:ds_algorithm} consists of estimating the halo density in spheres of radius $R_s$ centred around random points, which is then used to calculate the density PDF and define the DS quintiles. The density PDF itself depends on cosmology, which is the main source of information used in methods such as counts-in-cells statistics \citep{Uhlemann2020}. We also expect DS to be sensitive to this information, as any changes in the density PDF will translate into changes in the average density in each quintile, $\overline{\Delta}(R_s)$, which then propagates into changes in the observed multipoles.

\begin{figure}
    \centering
    \includegraphics[width=\columnwidth]{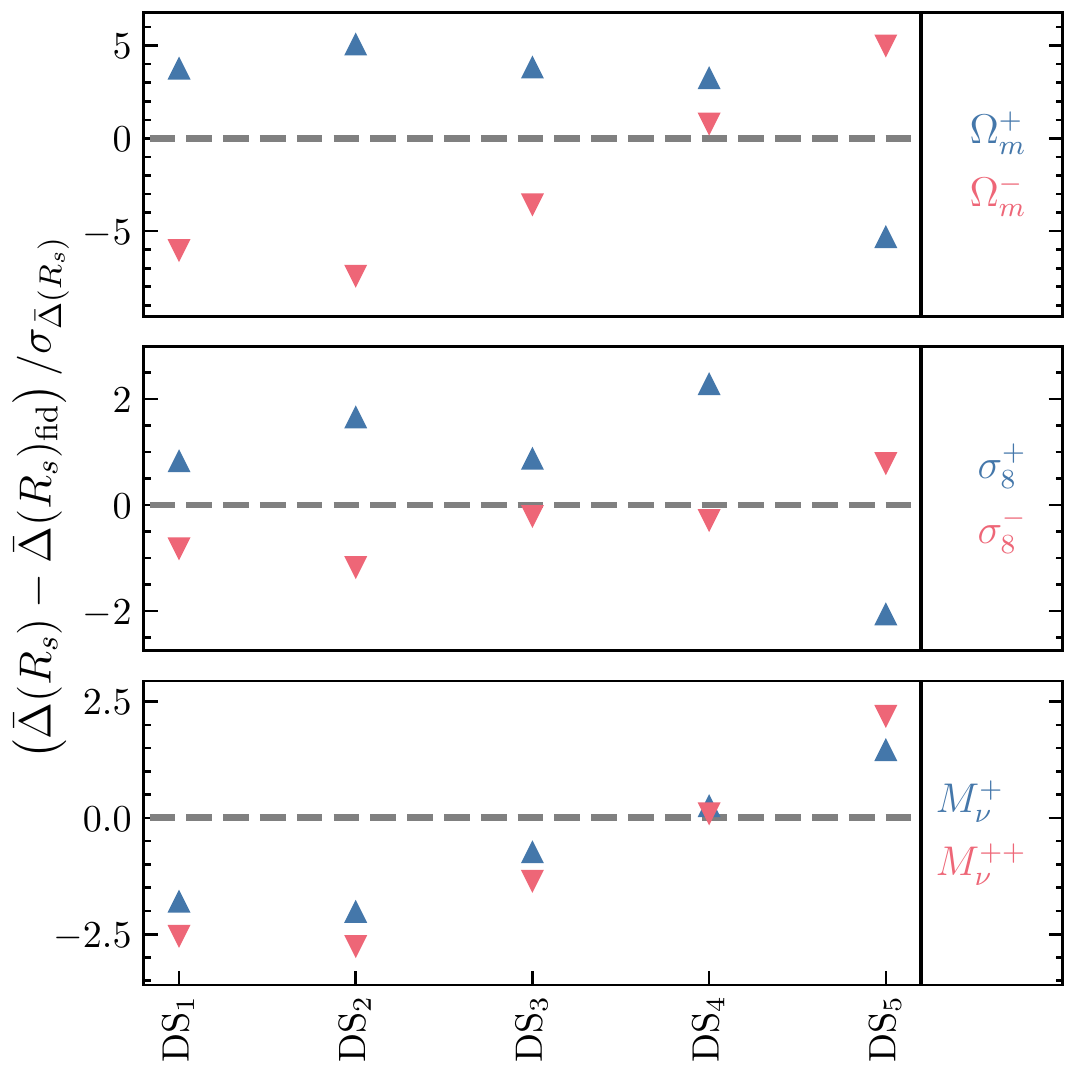}
    \caption{Response of the average density of each DS quintile to changes in cosmology. The vertical axis shows the difference in average density that is produced when we change $\Omega_{\rm m}$, $\sigma_8$ or $M_\nu$, in units of the 1-$\Sigma$ errors of the density. The horizontal axis shows results for each quintile separately. \href{https://github.com/epaillas/densitysplit-fisher/blob/master/Figure7.py}{\faFileCodeO}}
    \label{fig:density_quintiles}
\end{figure}

Figure \ref{fig:density_quintiles} illustrates this by showing how the average density per quintile responds to changes in the cosmological parameters. Increasing $\Omega_{\rm m}$ makes ${\rm DS}_1$, ${\rm DS}_2$, ${\rm DS}_3$, and ${\rm DS}_4$ denser, while the opposite happens for ${\rm DS}_5$.  On the one hand, given that we have fixed the minimum halo mass, increasing $\Omega_{\rm m}$ will increase the number of halos above this threshold. For the densest quintile, ${\rm DS}_5$, the increased merger rate could reduce the number of halos in a given sphere. On the other hand, when all other parameters are kept fixed, the effect of raising $\Omega_{\rm m}$ is to reduce the amplitude of the galaxy or halo power spectrum \citep{Kobayashi2020} by reducing the halo bias with respect to the underlying matter distribution, which brings the density of the quintiles slightly closer to the cosmic average. Changing $\sigma_8$ produces a similar effect on the quintiles, which is again related to an increase in the number of halos above the mass threshold and a reduced halo power spectrum for larger $\sigma_8$ values.

The effect of varying the neutrino mass goes in the opposite direction. Having a non-zero neutrino mass lowers the density from ${\rm DS}_1$ to ${\rm DS}_3$ and boosts the density in ${\rm DS}_5$.  This effect is very similar to that of decreasing $\Omega_{\rm m}$, since increasing the mass of neutrinos reduces the amount of cold dark matter. This is consistent with the picture that neutrinos, which do not cluster below their free-streaming scale, reduce the growth of cold dark matter perturbations. Although massive haloes can still form in the peaks of the density field and be resolved in Quijote, haloes forming in shallower regions of the density field will not reach masses above our selection threshold. The overall effect is an increased halo bias with respect to the fiducial case with $M_\nu = 0$ \citep{Kreisch2019}, which in turn makes the voids emptier and the clusters denser.

\subsection{Comparing the information content of density-split clustering to two-point statistics} 
\label{subsec:ds_vs_tpcf}

\begin{figure*}
    \centering
    \includegraphics[width=0.8\textwidth]{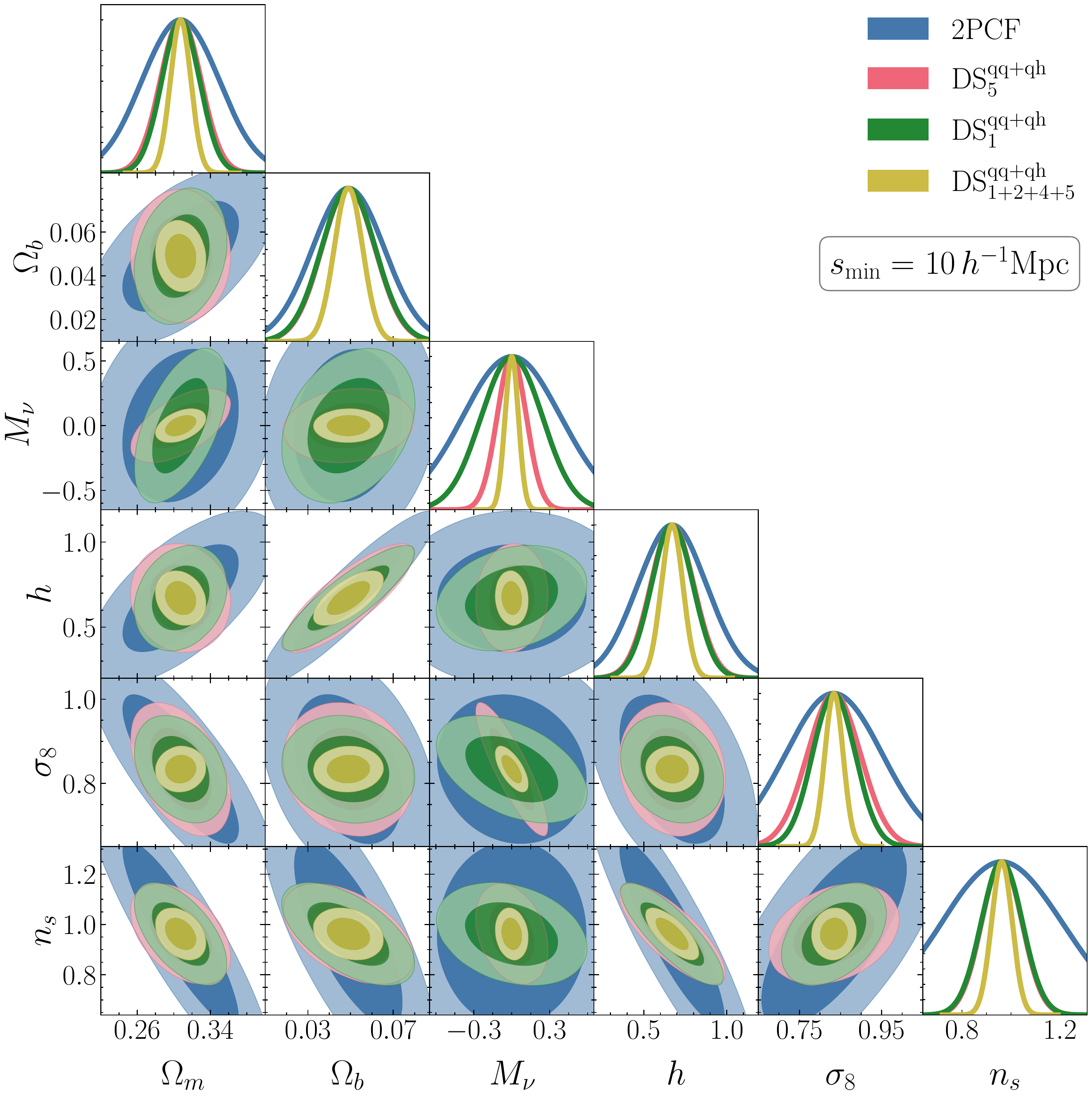}
    \caption{Fisher forecasts for constraints on the $\nu \Lambda$CDM model parameters from the 2PCF multipoles (blue) and density-split clustering using only voids (${\rm DS_1}$, green), clusters (${\rm DS_5}$, red) or the combination of all quintiles (yellow). \href{https://github.com/epaillas/densitysplit-fisher/blob/master/Figure8.py}{\faFileCodeO}}
    \label{fig:likelihood_full}
\end{figure*}

\begin{figure}
    \centering
    \includegraphics[width=\columnwidth]{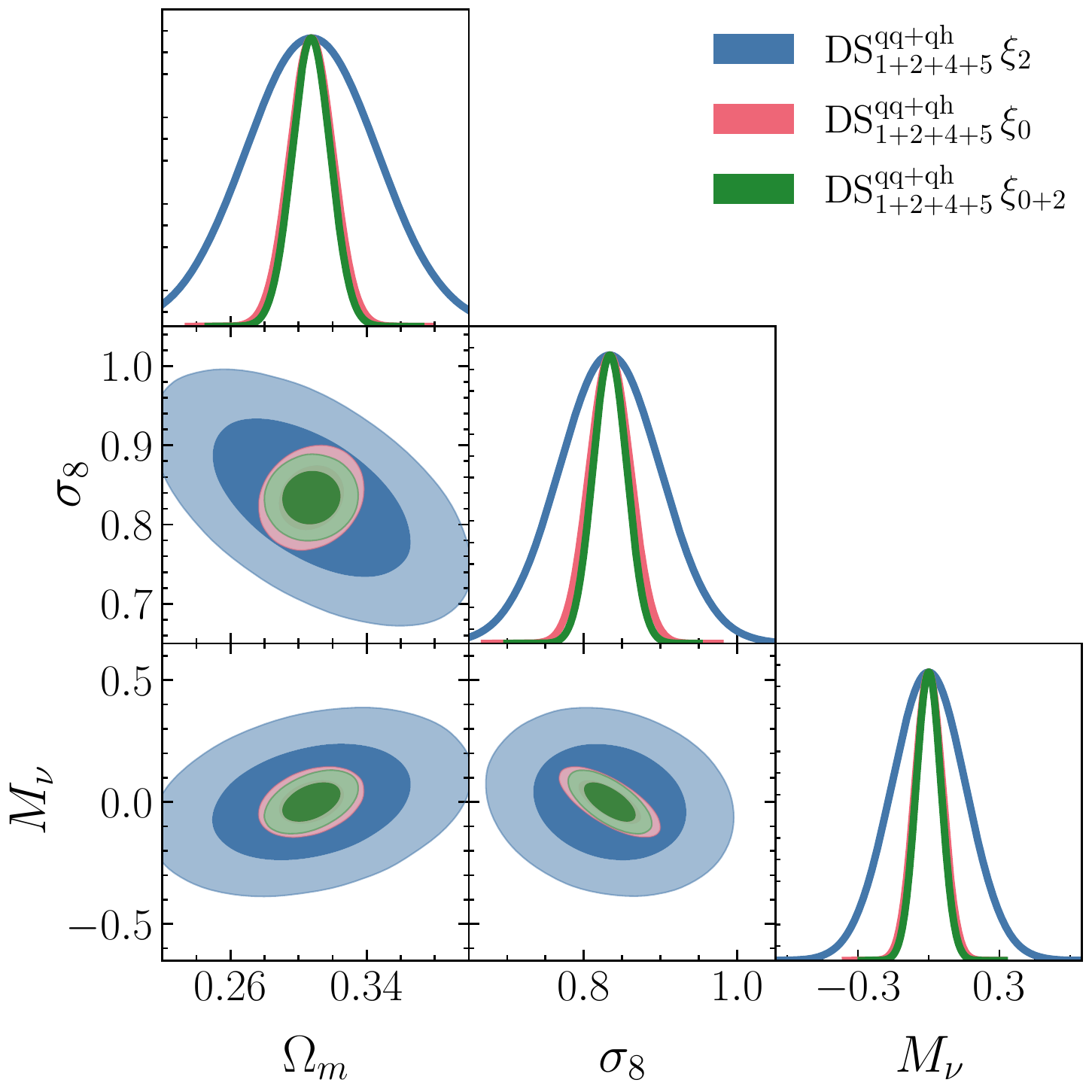}
    \caption{Comparison of the cosmological constraints from density-split clustering when using only the quadrupole (blue), monopole (red) or the combination of the two (red). \href{https://github.com/epaillas/densitysplit-fisher/blob/master/Figure9.py}{\faFileCodeO}}
    \label{fig:likelihood_mono_quad}
\end{figure}

In this section, we present the constraints obtained on the cosmological parameters through Eq.~(\ref{eq:cramer_rao}) and Eq.~(\ref{eq:fisher}). Unless stated otherwise, the DS constraints we show correspond to the $z$-split scenario, i.e., when density quintiles are defined in terms of the redshift-space overdensities.

Modelling the full cosmological dependence of the real-space or redshift-space-identified quintiles analytically would be challenging. In fact, previous studies \citep{Paillas2021} have only modelled the real-to-redshift space mapping. However, the Fisher formalism allows us to estimate the entire information content from direct measurements in N-body simulations.

In Fig.~\ref{fig:likelihood_full} we compare the constraints obtained by combining the DS autocorrelation and cross-correlation functions of four quintiles, ${\rm DS}_{1+2+4+5}^{\rm qq+qh}$, against the halo 2PCF, using multipoles within the scale range\footnote{We limit the measurements to scales larger than $10\, h^{-1}{\rm Mpc}$ since we are only analysing central halos, whose behaviour will be very different from that of galaxies on small scales, and because on these scales the effects of baryonic physics would be negligible.} $10 < s < 150\, {h^{-1}{\rm Mpc}}$. We can observe how DS can break some key parameter degeneracies that result when analysing two-point statistics, such as the one between $\Omega_{\rm m}$ and $\sigma_8$, or that of $n_s$ and $\sigma_8$. In particular, when we combine the information from all quintiles, the degeneracy between $M_\nu$ and the other parameters is significantly reduced. The standard halo 2PCF suffers from the well-known degeneracy found between $\sigma_8$ and $M_\nu$, which limits its constraining power. Although the individual quintiles ${\rm DS}_{1}^{\rm qq+qh}$ and ${\rm DS}_{5}^{\rm qq+qh}$ also exhibit this degeneracy to some extent, the combined DS dataset is able to reduce it due to the different sensitivity of each density environment to these parameters. Overall, ${\rm DS}_{1+2+4+5}^{\rm qq+qh}$ increases the constraining power with respect to the halo 2PCF by a factor of approximately $4$,  $7$, $3$, $3$, $6$, and $5$ for $\Omega_{\rm m}$, $M_{\nu}$, $\Omega_{\rm b}$, $h$, $n_s$, and $\sigma_8$, respectively.

The noisy derivatives of the quadrupole for certain quintiles shown in Fig.~\ref{fig:derivatives} might raise a concern about the robustness of the estimation of the information content of DS in the analysis. To assess this, in Fig.~\ref{fig:likelihood_mono_quad} we show constraints obtained by only fitting the monopole or the quadrupole moments of the correlation functions. We find that most of the constraining power is actually coming from the monopole alone (which has a higher signal-to-noise and thus smoother numerical derivatives), while the quadrupole only adds a marginal contribution to the combined power. Although we only show constraints for ${\rm \Omega_{\rm m}}$, $\sigma_8$, and ${\rm M_{\nu}}$, we have verified that the same trend is present in other regions of the parameter space.

\begin{figure}
    \centering
    \includegraphics[width=\columnwidth]{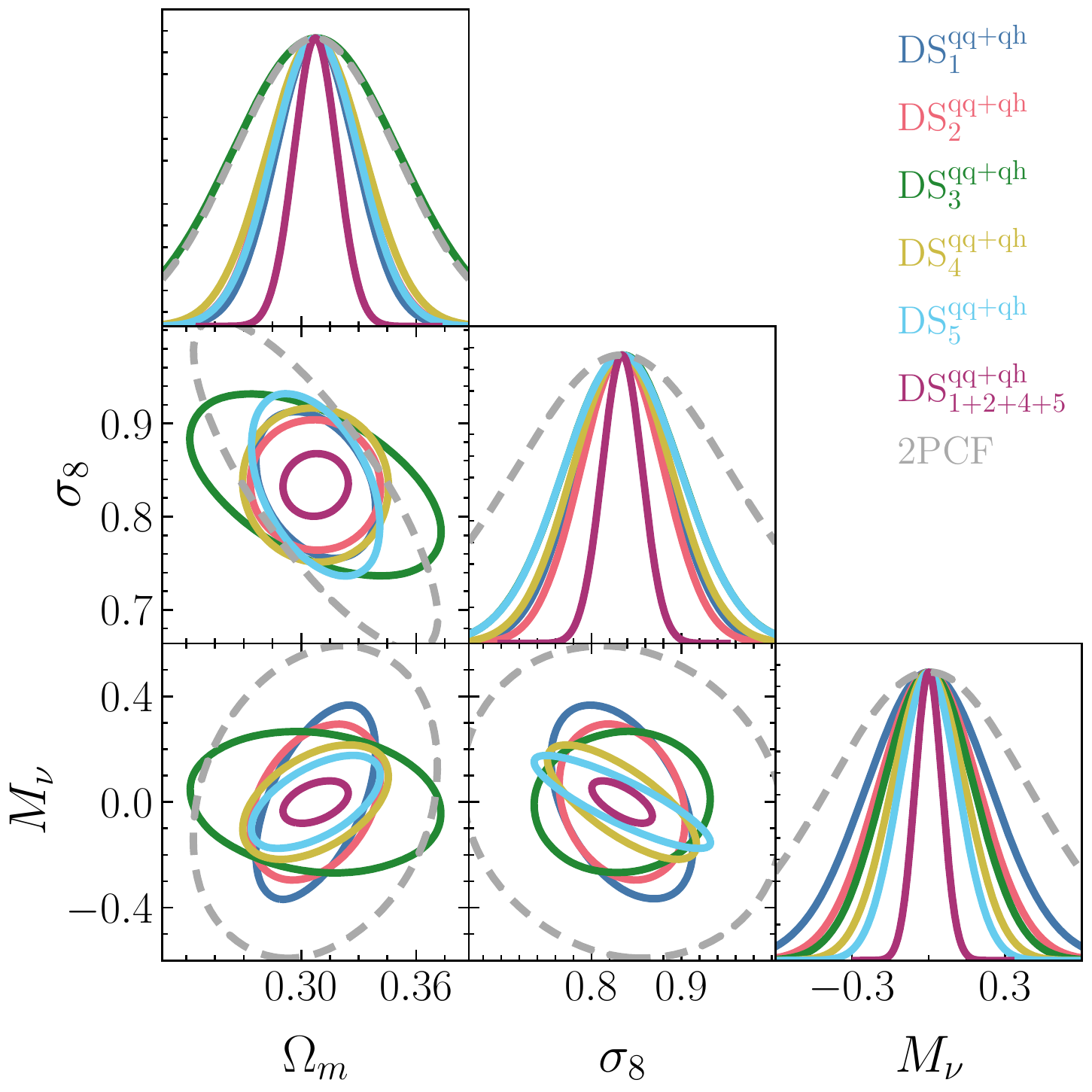}
    \caption{Constraints on the cosmological parameters from individual and combined DS quintiles identified in redshift space (solid). The constraints from the two-point correlation function are shown by the grey, dashed contours for comparison. \href{https://github.com/epaillas/densitysplit-fisher/blob/master/Figure10.py}{\faFileCodeO}}
    \label{fig:likelihood}
\end{figure}

In Fig.~\ref{fig:likelihood} we show the individual contribution of each quintile to the parameter constraints. Interestingly, we find that ${\rm DS}_{1}$ produces the weakest constraints for the sum of neutrino masses after marginalising over all other parameters. On the other hand, as we have explicitly checked, it produces the tightest constraints when all other parameters are fixed. One expects underdense regions to be more sensitive to the properties of neutrinos since their free-streaming motions imply that the ratio of neutrino density to that of dark matter is higher in void regions than in overdensities. However, degeneracies between the different cosmological parameters degrade the constraining power of underdense regions in DS. Furthermore, most quintiles individually produce tighter constraints than the 2PCF, except for ${\rm DS}_{3}$ and $\Omega_{\rm m}$.

\begin{figure}
    \centering
    \includegraphics[width=\columnwidth]{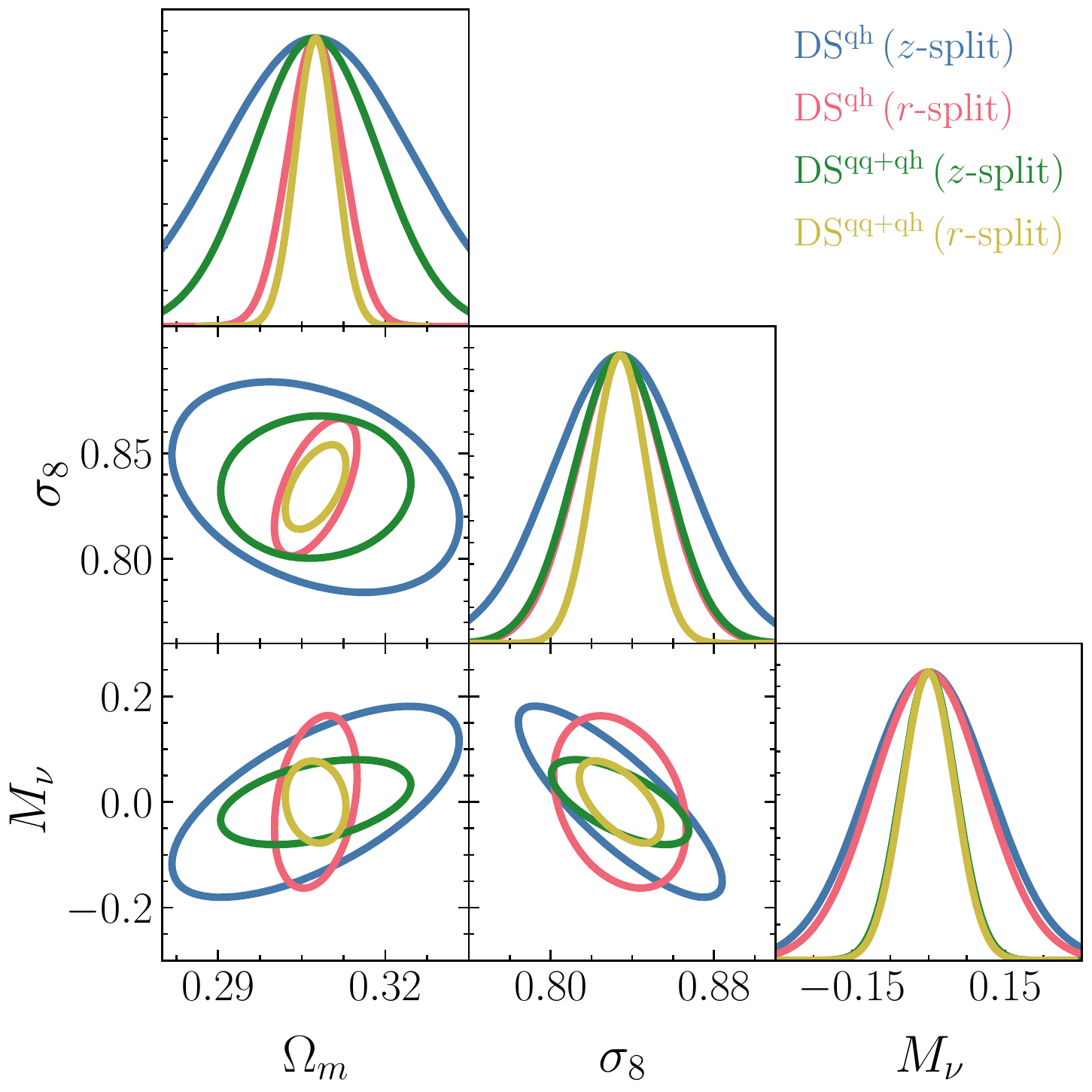}
    \caption{We compare the constraints obtained through cross-correlations of the density-split centres and the entire halo field, ${\rm DS}^{\rm qh}$, to the those obtained from the combination of cross-correlations and autocorrelations of the density-split centres, ${\rm DS}^{\rm qq+qh}$. We show both results for density-split centres identified in real space ($r$-split), and density-split centres identified in redshift space ($z$-split). This figure demonstrates that quintile autocorrelations, ${\rm DS}^{\rm qq}$, have a bigger impact in redshift identified quintiles than they do in real identified ones. \href{https://github.com/epaillas/densitysplit-fisher/blob/master/Figure11.py}{\faFileCodeO}}

    \label{fig:likelihood_bridge}
\end{figure}

Figure \ref{fig:likelihood_bridge} compares the information content of DS clustering when the overdensities are identified in redshift ($z$-split) or real space ($r$-split). The combined constraints on the cosmological parameters are shown in Table~\ref{tab:constraints}. The real-space identification of quintiles consistently produces better parameter constraints, especially for the parameters $\Omega_{\rm m}$ and $\sigma_8$. When quintiles are identified in redshift space, some cosmological information is lost by the blurring of the DS quintiles. However, some of this lost information can be recovered through the quadrupole of autocorrelations when these are identified in redshift space. This can be seen in Fig.~\ref{fig:likelihood_bridge}: although the additional information contained in the autocorrelations is small for the $r$-split scenario, it has a large impact on improving the constraints for DS centres identified in redshift space. 

\subsection{Where does the additional information come from?}

\begin{figure*}
    \centering
    \begin{tabular}{ccc}
      \includegraphics[width=0.28\textwidth]{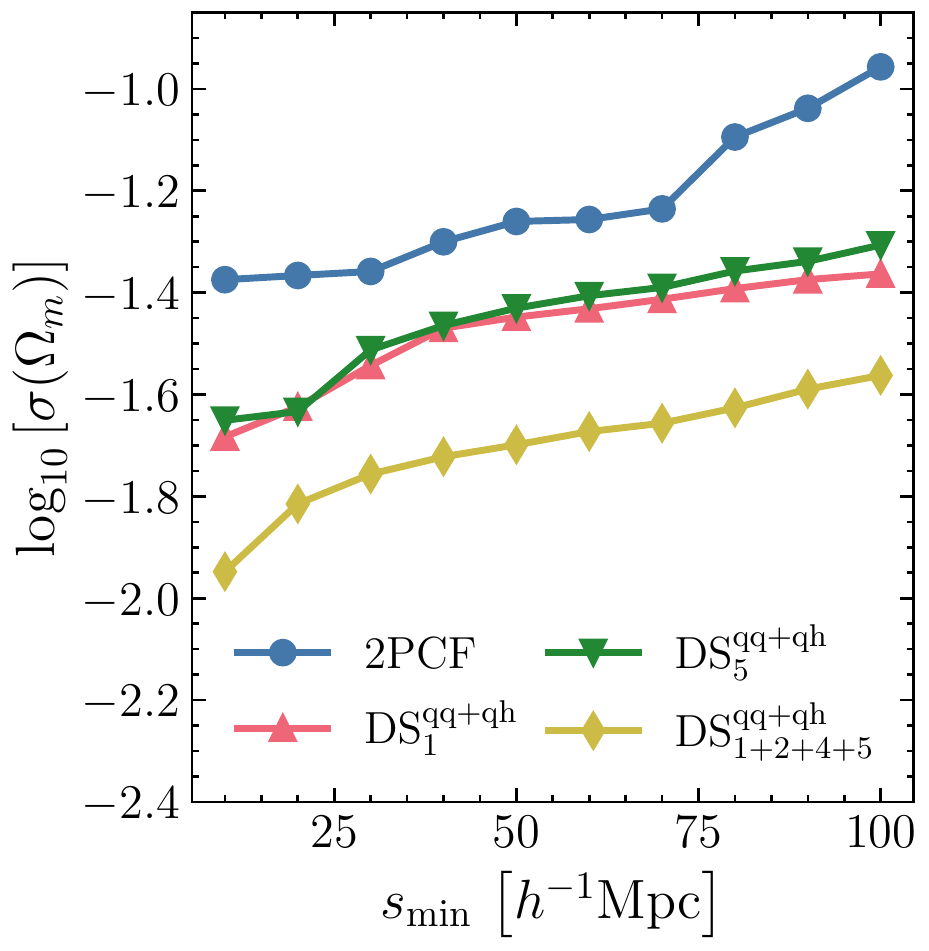} & \includegraphics[width=0.28\textwidth]{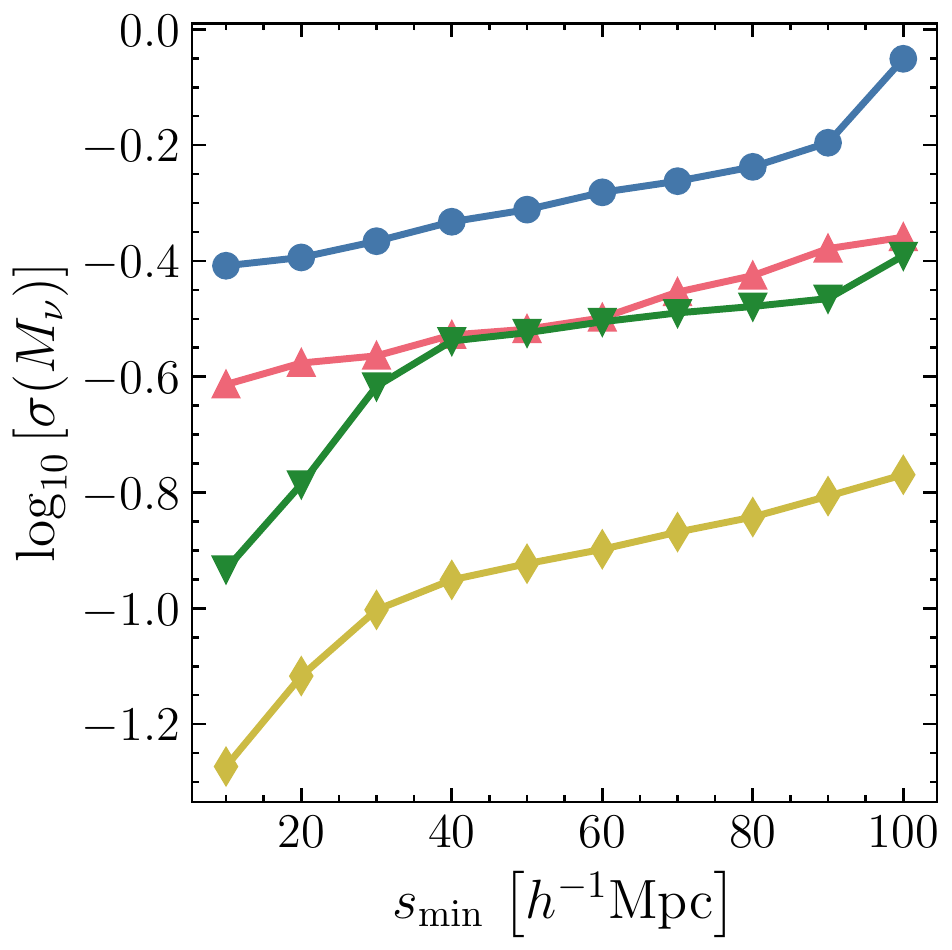} &
      \includegraphics[width=0.28\textwidth]{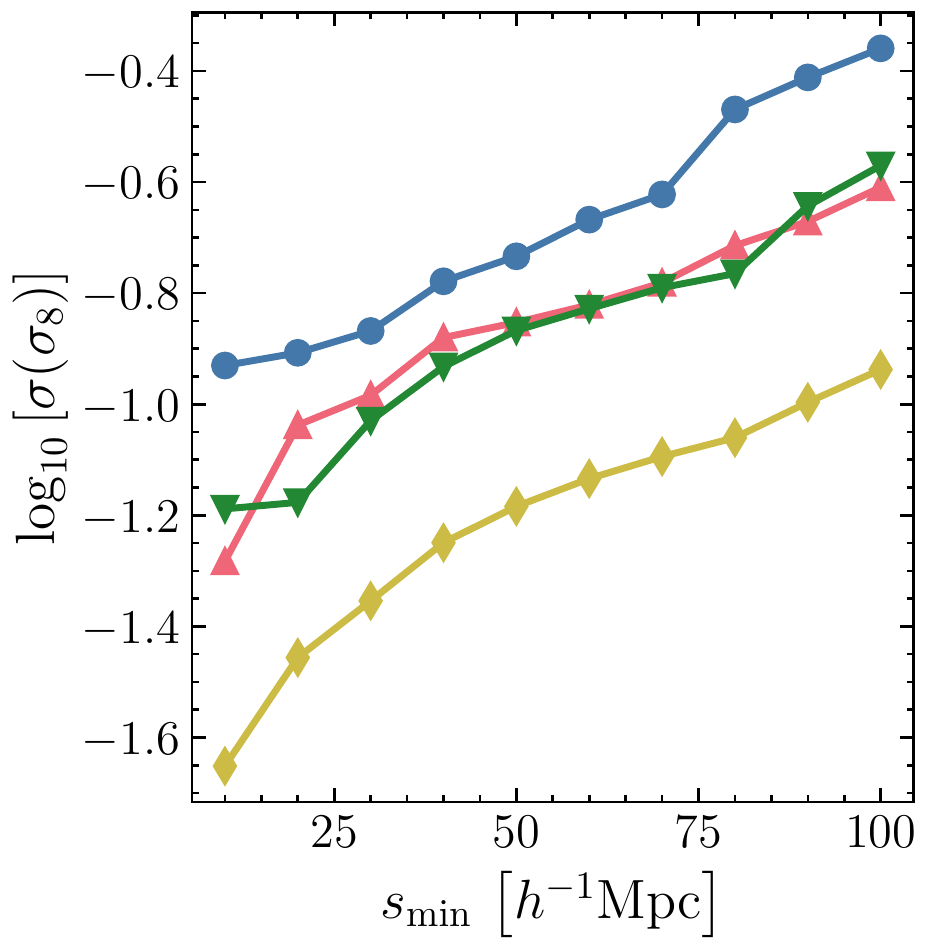} \\
      \includegraphics[width=0.28\textwidth]{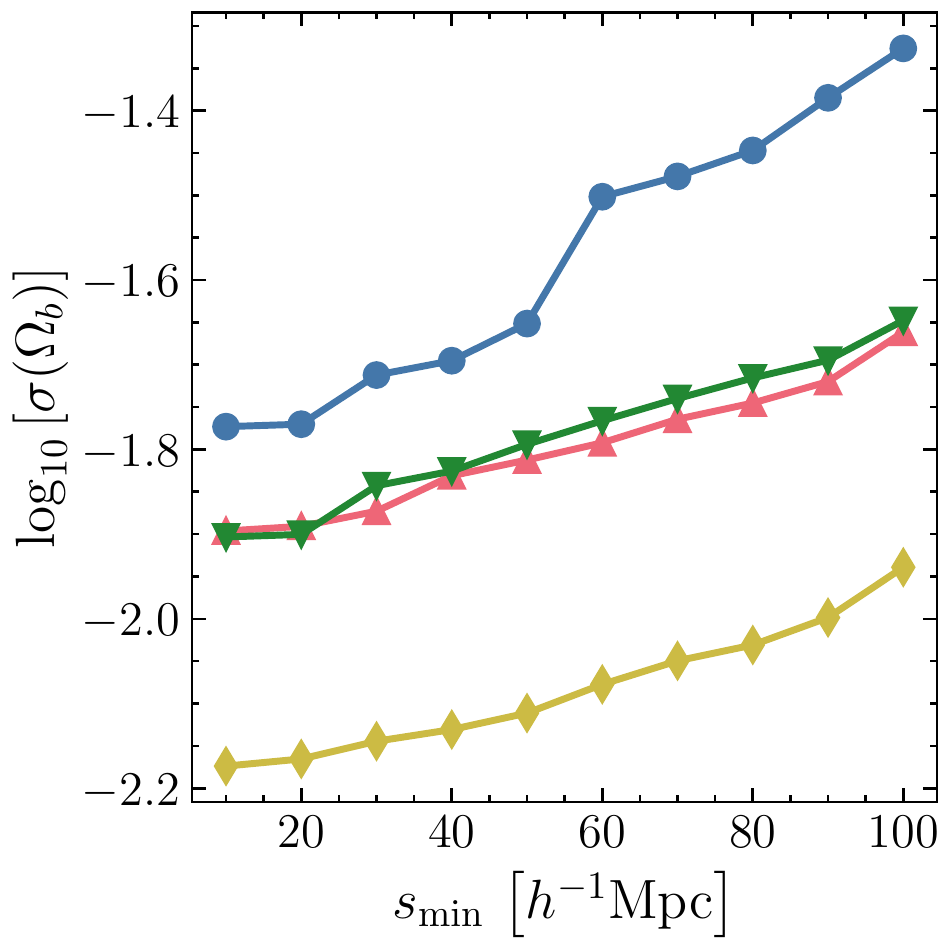} &
      \includegraphics[width=0.28\textwidth]{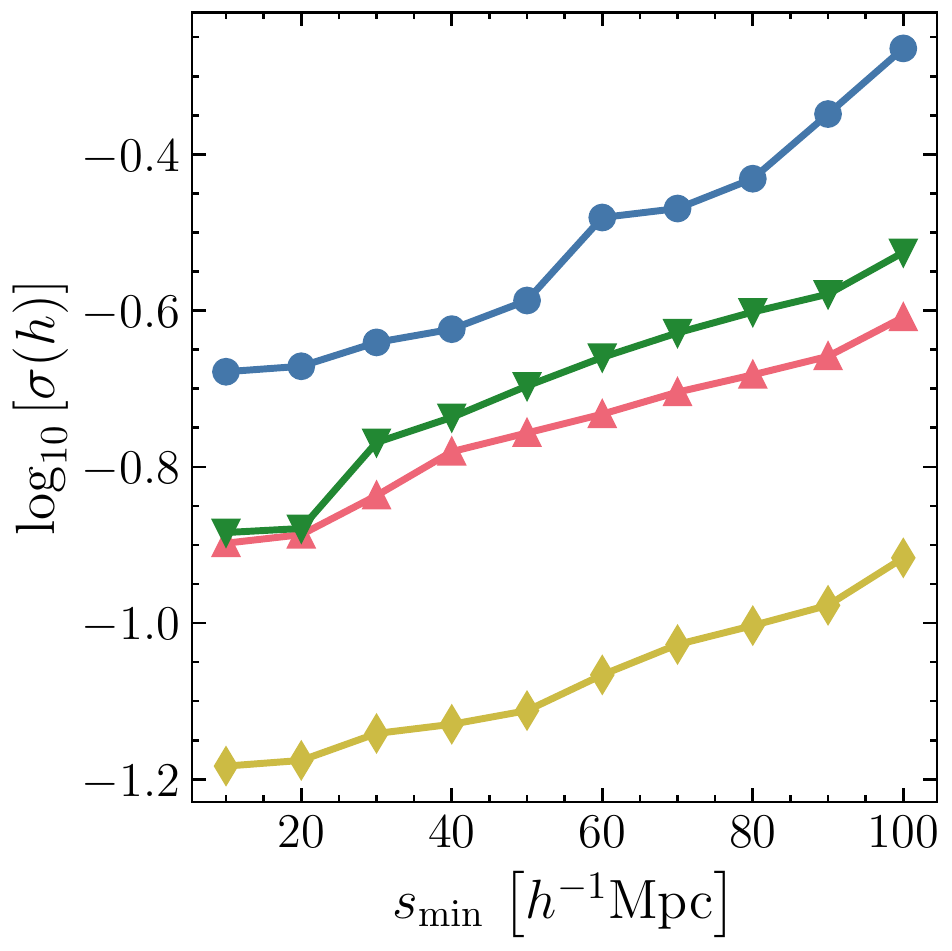}  & \includegraphics[width=0.28\textwidth]{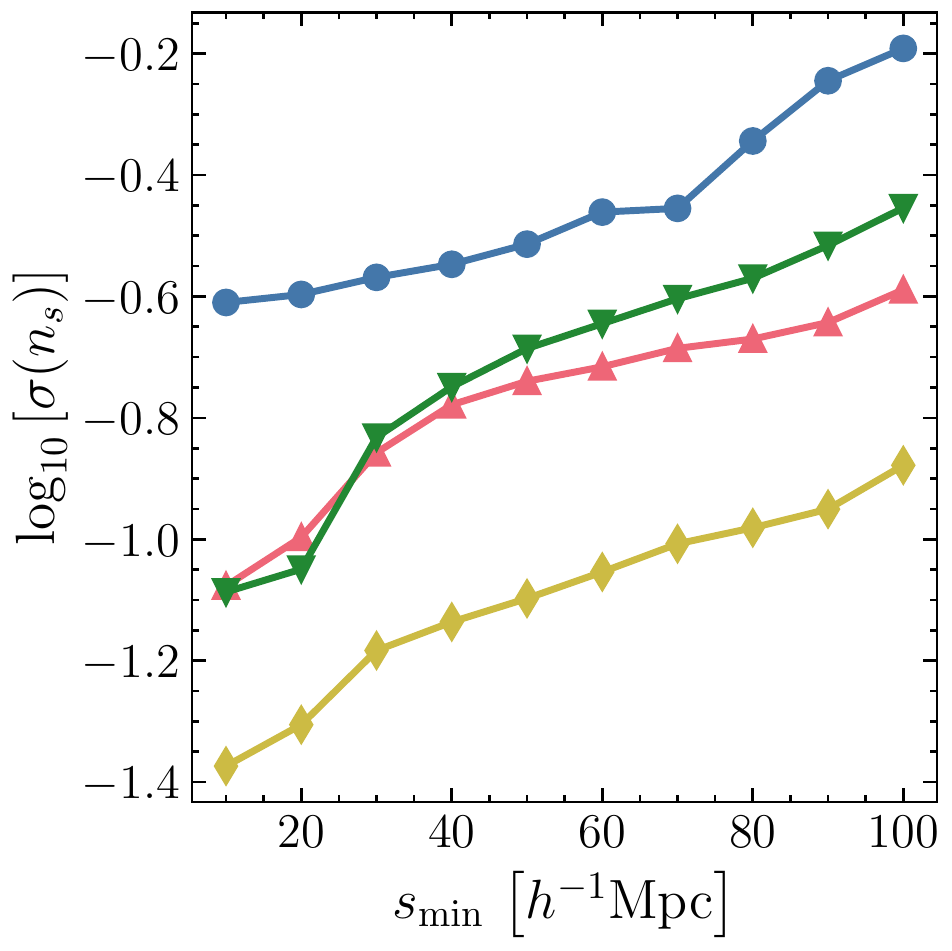}
    \end{tabular}
    \caption{Constraints on the cosmological parameters from DS and the 2PCF, as a function of the minimum scale used to calculate the Fisher matrix. We also include the individual constraints obtained through the two extreme quintiles, ${\rm DS_1}$ and ${\rm DS_5}$. \href{https://github.com/epaillas/densitysplit-fisher/blob/master/Figure12.py}{\faFileCodeO}}
    \label{fig:errors_scale}
\end{figure*}

We have seen that at the same fixed minimal scale, DS always outperforms 2PCF for constraining cosmological parameters. This remains true when we increase the minimal scale, as shown in Fig.~\ref{fig:errors_scale}. We can see that even when $s_{\rm min}$ is very large, e.g. $\sim 100\, h^{-1}{\rm Mpc}$, where we expect the density field to be close to Gaussian, the constraints from DS are still significantly tighter than 2PCF.  

There are at least two different effects that can lead to a better performance of DS over the 2PCF on different scales. First, DS is able to extract non-Gaussian features in the density field that are not fully captured by the 2PCF. This effect is expected to be more important for smaller $s_{\rm min}$ values, where stronger deviations from Gaussianity are found. Second, DS quintiles are defined in terms of the density contrasts in spheres with the radius $R_s=20\,h^{-1}{\rm Mpc}$, so even when we truncate the multipoles at large $s_{\rm min}$ values, DS still carries information about the PDF of the density field smoothed at $R_s$, which is not present in the 2PCF multipoles.

To double check the above reasoning, we test it with ideal Gaussian random fields. Starting from primordial power spectra with the same parameters as those described in Table~\ref{tab:simulations}, we use \textsc{mockfactory}\footnote{\url{https://github.com/cosmodesi/mockfactory}} to generate a Gaussian random field at $z = 0.0$, sampled with particles with tracer bias similar to the Quijote haloes. We compute the 2PCF and DS correlation functions using 30 radial bins in the scale range $0 < r < 150\, h^{-1}{\rm Mpc}$ and estimate the Fisher matrix numerically as described in Sect.~\ref{sec:fisher_formalism}. For simplicity, all measurements are performed in real space, so that all information is contained in the monopole moment of the correlation functions.

\begin{figure*}
    \centering
    \begin{tabular}{cc}
      \includegraphics[width=0.45\textwidth]{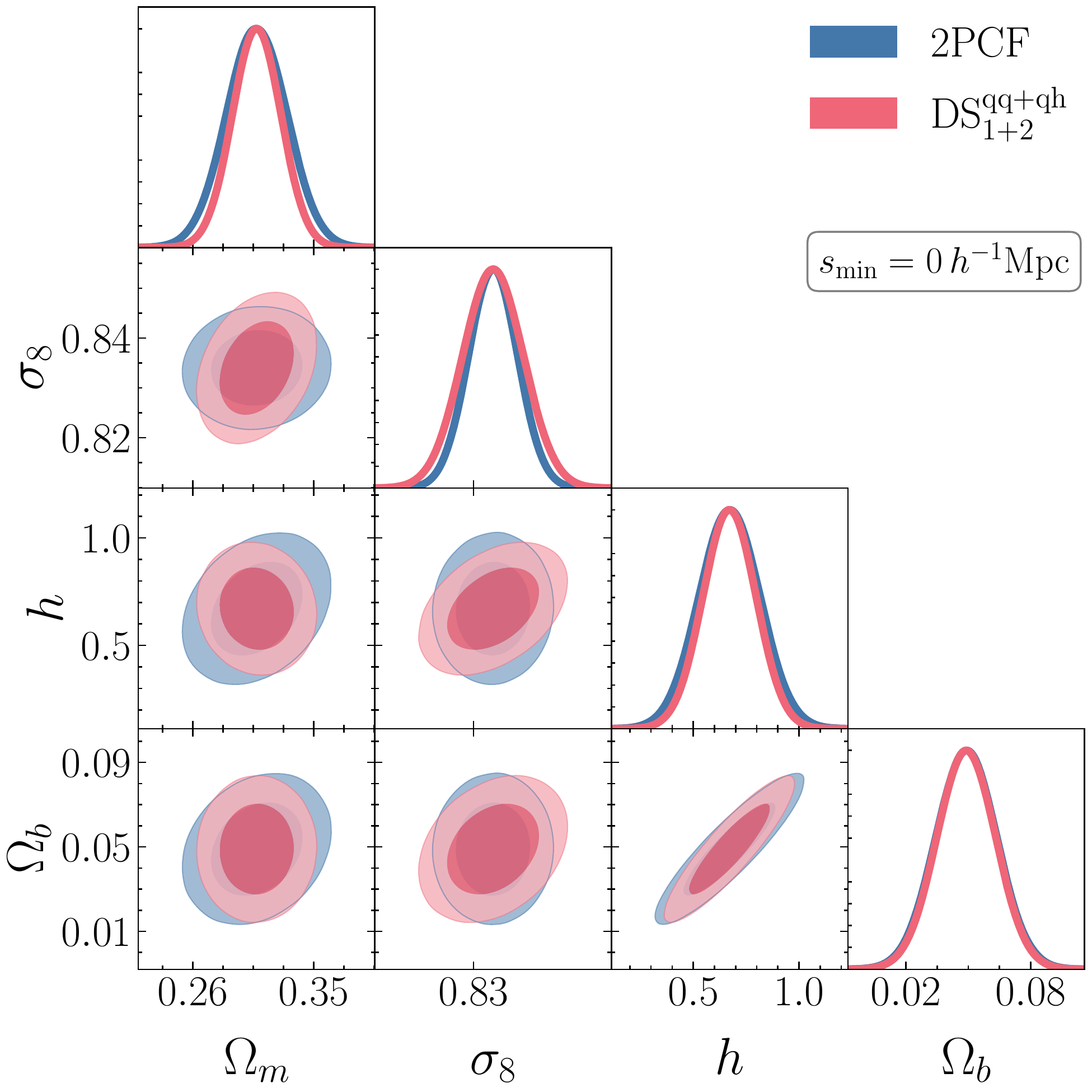}  & \includegraphics[width=0.45\textwidth]{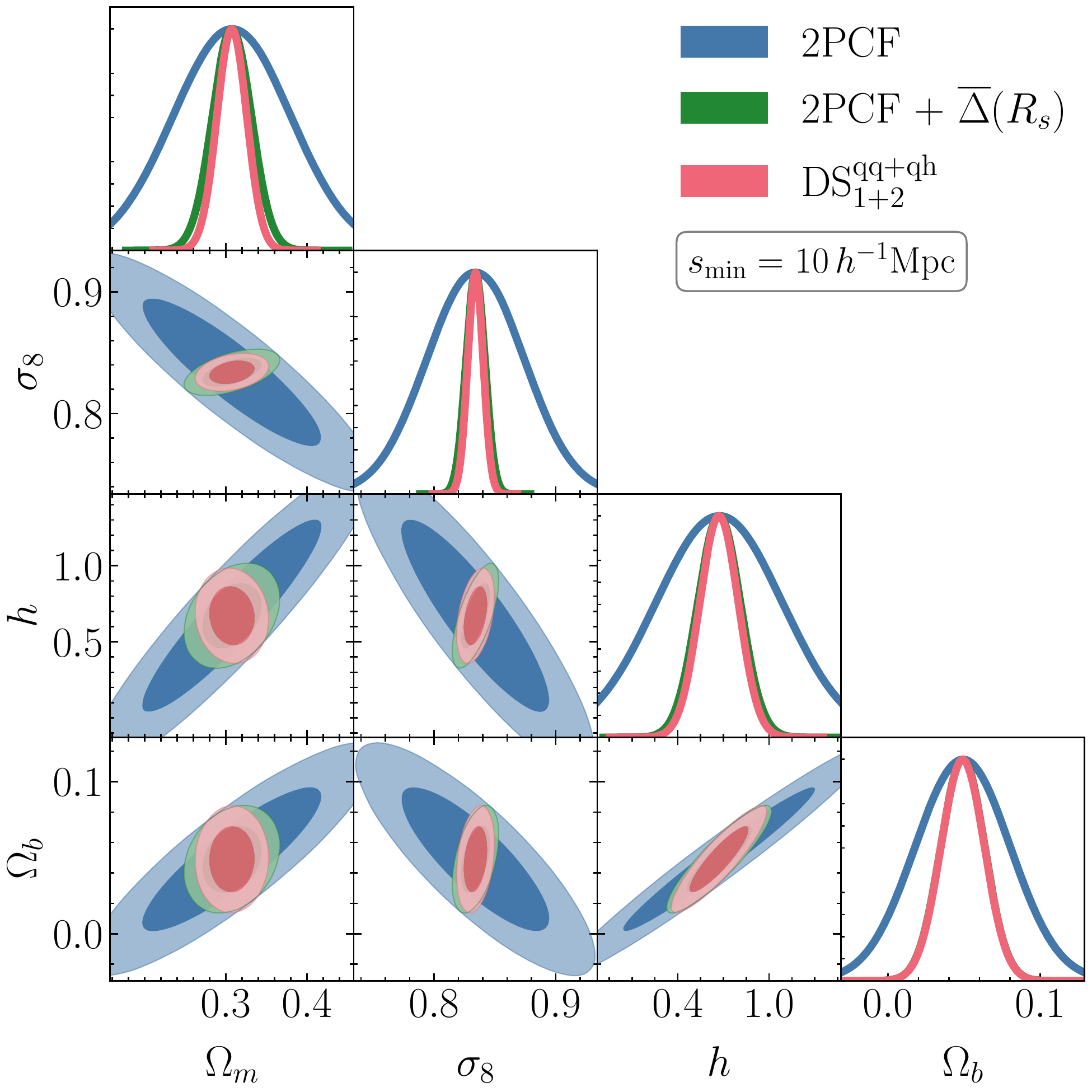}
    \end{tabular}
    \caption{Cosmological parameter constraints from DS and the 2PCF in a Gaussian random field, using scales down to $0\,h^{-1}{\rm Mpc}$ (left) or $10\,h^{-1}{\rm Mpc}$ (right) Blue: constraints from the real-space halo 2PCF. Red: constraints from the combination of DS cross-correlation and autocorrelation functions in real space. Green: constraints from the combination of the halo 2PCF and the average density in DS quintiles. We note that DS-related quantities only make use of the first two quintiles, ${\rm DS}_1$ \& ${\rm DS}_2$, which contain all the information if the density PDF is symmetric. \href{https://github.com/epaillas/densitysplit-fisher/blob/master/Figure13a.py}{\faFileCodeO} \href{https://github.com/epaillas/densitysplit-fisher/blob/master/Figure13b.py}{\faFileCodeO}}
    \label{fig:likelihood_gaussian}
\end{figure*}

In this Gaussian case, the 2PCF, which is a measure of the variance of the field as a function of scale, should be able to fully describe its statistical properties, and we expect DS and the 2PCF to contain the same cosmological information. We can see that this is indeed the case, as shown in the left-hand panel of Fig.~\ref{fig:likelihood_gaussian}. Under this setup, DS and the 2PCF show similar constraints on $\Omega_{\rm m}$, $\sigma_8$, $h$, and $\Omega_{\rm b}$ using the full-scale range.\footnote{While the 2PCF almost perfectly matches the DS constraints for $\Omega_{\rm b}$ and $h$, and it outperforms DS for $\sigma_8$, we find that DS yields constraints that are a factor of 1.2 better for $\Omega_{\rm m}$. Some of this discrepancy could be attributed to numerical errors in the Fisher matrix due to the finite number of mocks from which the numerical derivatives are estimated, although we have checked that the constraints converge to better than 10 per cent for the number of mocks that we used.}

The right-hand panel of Fig.~\ref{fig:likelihood_gaussian} repeats this comparison using a minimum scale $s_{\rm min} = 10\, h^{-1}{\rm Mpc}$. In this case, DS leads to significantly improved constraints over the 2PCF for all parameters. 
This may go against the intuition that DS should not be able to outperform the 2PCF in the Gaussian scenario. However, as discussed in the beginning of this subsection, we should keep in mind that the DS quintiles are defined in terms of the halo densities in spheres of radius $R_s = 20\,h^{-1}{\rm Mpc}$. This makes the DS quintiles sensitive to the variance of the field within $R_s$, even when the multipoles are truncated at $s_{\rm min} = 10\, h^{-1}{\rm Mpc}$ (as formally shown in Pinon et al. (in preparation)). To account for this effect, we include the average density in each quintile, $\overline{\Delta}(R_s)$, as part of the observable, calculating the Fisher matrix of the concatenated data vector 2PCF + $\overline{\Delta}(R_s)$, which accounts for the covariance between the two measurements. It can be seen from the figure that the resulting constraints from this combination match the constraints from DS much better, recovering the agreement seen earlier in the left-hand panel.

We note that in simulations where the density field is non-Gaussian (Quijote), we have explicitly checked that DS outperforms 2PCF + $\overline{\Delta}(R_s)$. This is because the addition of the $\overline{\Delta}(R_s)$ information is equivalent to sampling the density PDF at a single scale, $R_s$, which captures only part of the non-Gaussian information. On the other hand, the DS-halo cross-correlation in each quintile is equivalent to measuring the average enclosed halo overdensity around those DS centres.\footnote{This follows since ${\rm DS}^{\rm qh}_x(s)$ represents the average halo overdensity at distance $s$ from the DS centres in quintile $x$, and the enclosed overdensity $\overline{\Delta}(s)$ for the quintile is simply an integral of this.} Thus measuring the cross-correlation ${\rm DS}^{\rm qh}_{1+2+4+5}(s)$ is equivalent to sampling the density PDF at a range of scales $s$.

In summary, the combination of the 5 DS-halo correlations measures the PDF of the density field as a function of scale i.e. the histograms of $\Delta(R)$. It thus captures non-Gaussianities at all scales of our measurements, and outperforms 2PCF for cosmological constraints. When there is a minimal scale cut off, DS can outperform 2PCF even more because it implicitly contains information about the PDF of the density at the smoothing scale.

We caution that the above reasoning may be incomplete and that there may be room for other reasons to account for the additional information in DS. We will have more discussions on this in Sect.~\ref{sec:discussion_conclusions}, and leave a more rigorous study on this point for a future work.

\subsection{Information content of reconstructed density-split multipoles}\label{sec:bias_reconstruction}

In the previous section, we showed that performing the density split on the real-space galaxy field in principle provides significantly more information than doing so in redshift space, as the Fisher information content of the $r$-split multipoles is higher. However, in practical applications to data, the real-space galaxy positions would not be available to allow such a measurement.

One way to proceed would be to accept the loss of information associated with the redshift-space density split procedure and to use the $z$-split multipoles alone for cosmological inference. (While we currently lack an analytical model to predict $r$-split or $z$-split multipoles from first principles, we envisage an inference procedure based on constructing an emulator for these quantities using $N$-body simulations; such an emulator could equally be constructed for either $r$-split or $z$-split quantities.)

\begin{figure}
    \centering
    \includegraphics[width=\columnwidth]{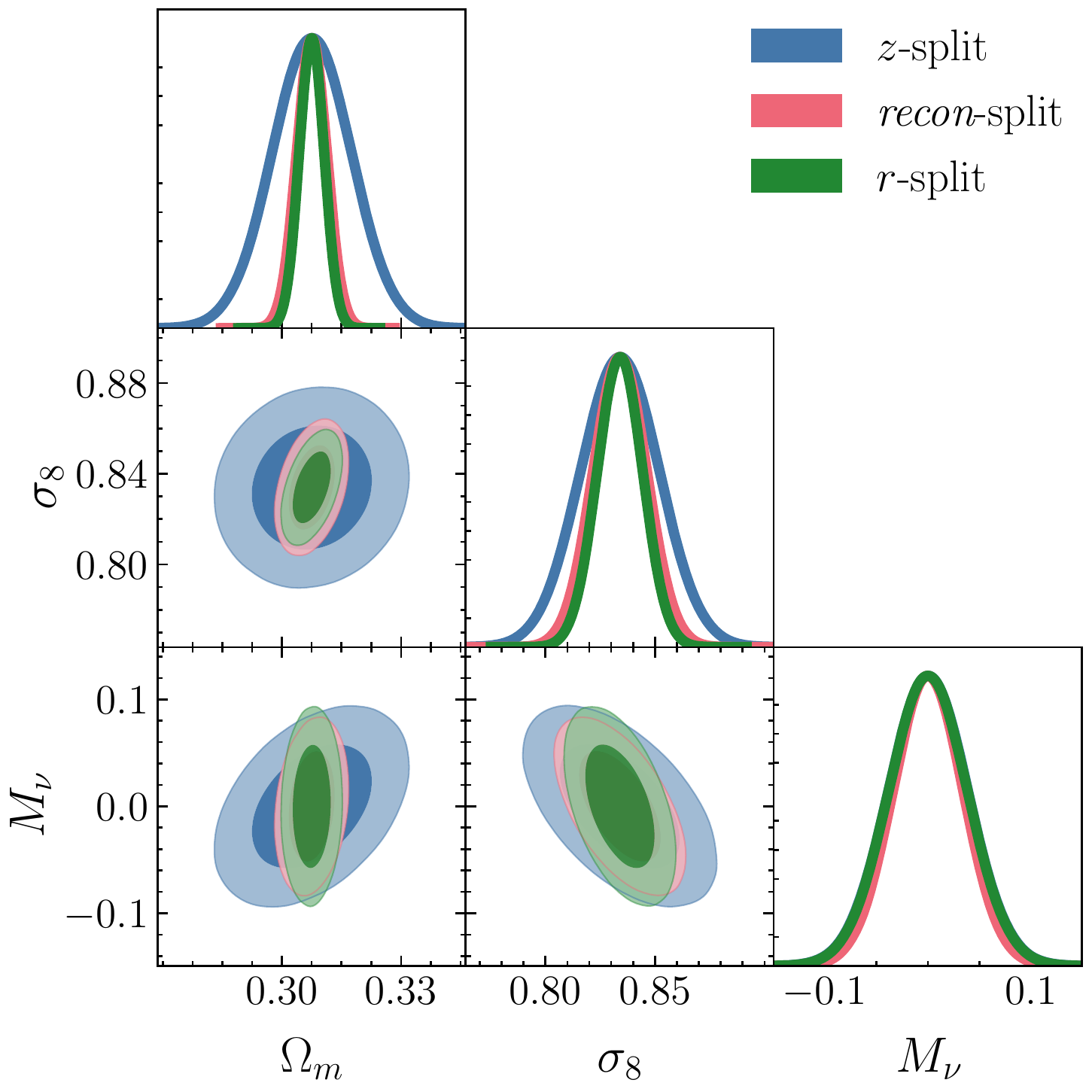}
    \caption{Marginalised constraints on $\Omega_{\rm m}$, $\sigma_8$ and $M_{\nu}$ when the density-split quintiles are identified in redshift space (blue), real space (green) or in a pseudo-real space where RSD has been removed using reconstruction (red). The constraints have been obtained by combining the cross-correlation and autocorrelation functions of all density quintiles, which is referred to as ${\rm DS}_{1+2+4+5}^{\rm qq + qh}$ in previous figures. \href{https://github.com/epaillas/densitysplit-fisher/blob/master/Figure14.py}{\faFileCodeO}}
    \label{fig:reconstruction_likelihood}
\end{figure}

On the other hand, in Sect.~\ref{sec:reconstruction} we also showed that it is possible to use a reconstruction method to recover approximate real-space galaxy positions before performing the density split, and that the ``\textit{recon}-split'' multipoles thus obtained closely match the $r$-split multipoles. Therefore, the use of \textit{recon}-split multipoles could, in principle, allow the recovery of much of the information contained in the $r$-split multipoles that is lost when using the $z$-split. This is shown in Fig.~\ref{fig:reconstruction_likelihood}, where we compare the marginalised contours of $\Omega_{\rm m}$, $\sigma_8$, and $M_{\nu}$ between the different DS identification scenarios. In terms of the information content, \textit{recon}-split largely outperforms $z$-split for $\Omega_{\rm m}$ and $\sigma_8$, resulting in constraints that are only a factor of 1.21 and 1.18 weaker than $r$-split, respectively. For $M_{\nu}$, \textit{recon}-split and $r$-split agree within 10 per cent, while \textit{recon}-split outperforms $z$-split by a factor of 1.13.

As practical reconstruction methods are not perfect, there are small differences between the $r$-split multipoles and those that can be obtained from the reconstruction procedure (Fig.~\ref{fig:reconstructed_multipoles}). Any cosmological analysis using the \textit{recon}-split multipoles would therefore require theoretical modelling that specifically accounts for these differences. Constructing an emulator for the \textit{recon}-split multipoles would be a more complicated proposition than doing so for $z$-split. Apart from the increased computational cost of reconstructing the density field before splitting the densities, we need to consider the additional dependence on cosmology due to the sensitivity of reconstruction to the ratio between the linear growth rate and the tracer bias $f/b$. The results might also be sensitive to the different choices of configuration parameters in the algorithm, such as the scale used to smooth the density field and the resolution of the grid used to perform the Fourier space operations. We plan to address the feasibility of modelling such effects within the DS framework in future work.

It might be tempting to avoid these difficulties by simply using model predictions for the $r$-split multipoles -- which would be easier to construct -- as a proxy for the \textit{recon}-split multipoles that are more practical to measure in survey data. However, in this case, the differences seen in Fig.~\ref{fig:reconstructed_multipoles} could potentially lead to systematic errors in the inferred cosmological parameters. In the remainder of this section, we investigate and quantify this possibility.

We estimate the bias in the inferred cosmological parameters introduced by the imperfections in reconstruction using the Fisher matrix \citep{Huterer2005}
\begin{align} \nonumber
\label{eq:biases}
\delta \theta_\alpha &= \langle \theta^{\mathrm{\textit{recon}}-{\rm split}}\rangle - \langle \theta^{\mathrm{\textit{r}}-{\rm split}} \rangle \\ &=\sum_\beta  \mathcal{F}_{\alpha \beta}^{-1} \sum_{ij} \left[ \bm d^{\mathrm{\textit{recon}}-{\rm split}}_i - {\bm d}^{\mathrm{\textit{r}}-{\rm split}}_i \right] C_{ij}^{-1} \frac{\partial {\bm d}^{\mathrm{\textit{r}}-{\rm split}}_j}{\partial \theta_\beta}
\end{align}

\begin{figure*}
    \centering
    \includegraphics[width=\textwidth]{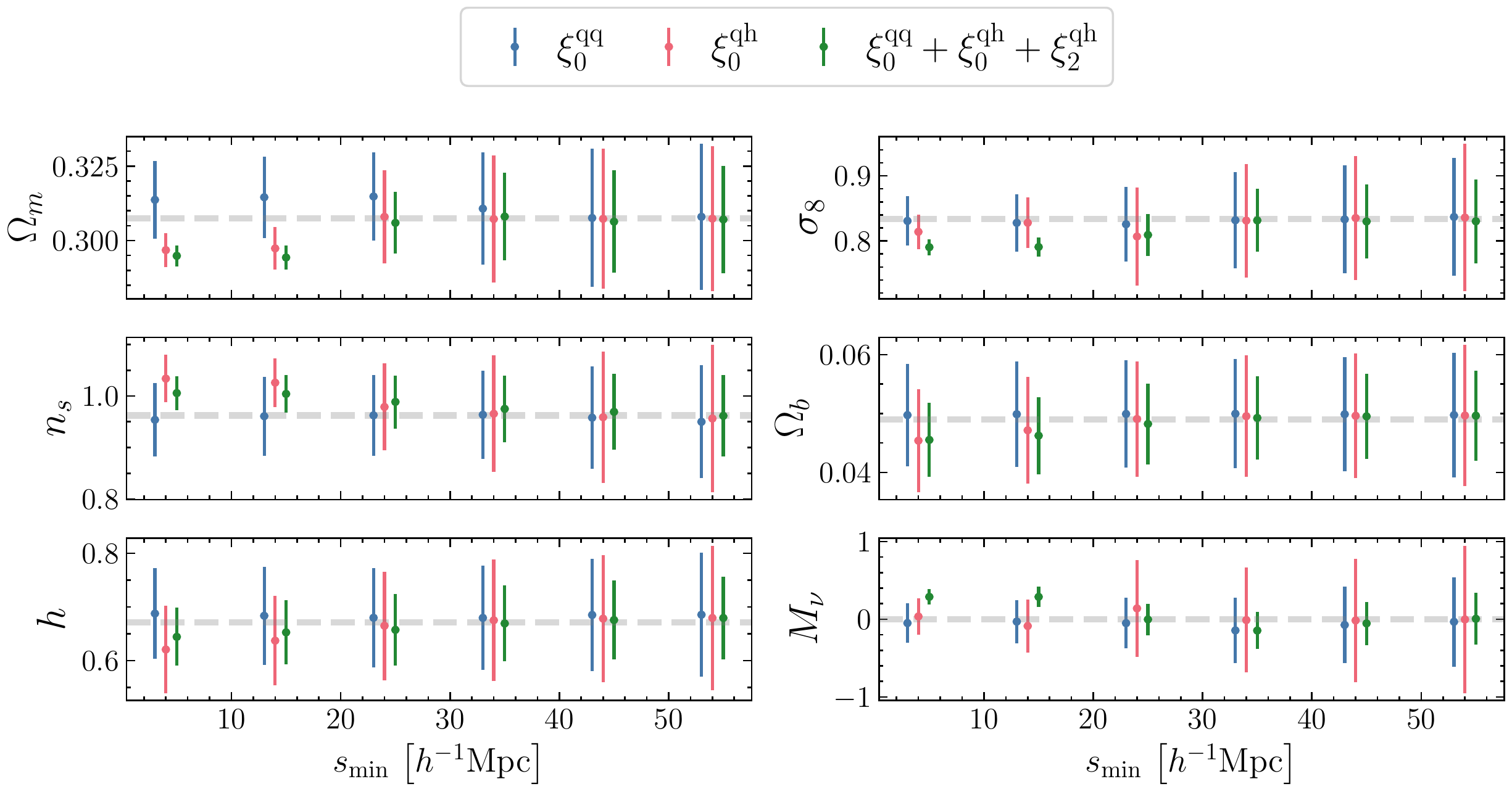}
    \caption{Bias in the cosmological parameters introduced by systematic errors caused by reconstructing the halo's real space positions, computed using Eq.~(\ref{eq:biases}). The true value of the parameters is shown on a gray dashed line. We show the bias introduced by each of the statistics used to infer the cosmological parameters: i) $\xi_0^\mathrm{qq}$, the monopole of the quintile autocorrelation, ii) $\xi_0^\mathrm{qh}$, the monopole of the cross-correlations between quintiles and halos, and iii) $\xi_0^\mathrm{qq}+\xi_0^\mathrm{qh}+\xi_2^\mathrm{qh}$, the combination of all the above with the quadrupole of the cross-correlations between quintiles and halos. The error bars show the statistical uncertainty associated to a $(1\,h^{-1}{\rm Gpc})^3$ volume. Points that are plotted next to each other were conducted at the same $s_{\rm min}$, but are horizontally shifted for clarity. \href{https://github.com/epaillas/densitysplit-fisher/blob/master/Figure15.py}{\faFileCodeO}}
    \label{fig:reconstruction_bias}
\end{figure*}

In Fig.~\ref{fig:reconstruction_bias}, we show the biases in the inferred cosmological parameters that would be caused by such a misapplied model, as a function of the minimum scale considered in the analysis. $M_{\nu}$, $\Omega_{\rm m}$ and $\sigma_8$ are the parameters that are most affected by errors due to the imperfect reconstruction of halo positions. In particular, biases are found when including the monopole and quadrupole of cross-correlations between quintiles and the halo field, $\xi_{0,2}^\mathrm{qh}$, on scales smaller than the DS smoothing radius. In Fig.~\ref{fig:reconstructed_multipoles}, we have shown that the errors introduced by reconstruction mostly affect the quadrupole of cross-correlations. Using only the monopole of quintile autocorrelations, $\xi_0^\mathrm{qq}$, one can obtain unbiased constraints on the cosmological parameters using the full range of scales. However, the constraining power of autocorrelations on small scales is smaller than that of cross-correlations with the halo field, and therefore we would lose more information than if we were to estimate the overdensity around random centres directly in redshift space.

We note that the results presented in this section apply to a particular choice of reconstruction algorithm, which has been described in Sect.~\ref{sec:reconstruction}. Other algorithms \citep[e.g., ][]{White2015:1504.03677, Wang2022:1912.03392v3} may lead to different parameter constraints, although a detailed comparison of different reconstruction techniques is beyond the scope of this manuscript. 

As described in Sect.~\ref{sec:reconstruction}, reconstruction also smooths the density field below a given scale $R_s^{\rm recon}$, which is a free parameter in the algorithm. In our analysis, this scale was set to $10\,h^{-1}{\rm Mpc}$. We do not expect reconstruction to work below $R_s^{\rm recon}$, where the clustering information has been washed out, and consequently, the removal of RSD may be inaccurate. Future surveys, such as DESI-BGS \citep{Zarrouk2022}, are expected to reach much higher tracer number densities than those probed by Quijote, and the range of scales at which reconstruction is reliable may differ. We plan to study this in further detail in future work.

In summary, the information content in the resulting \textit{recon}-split multipoles is similar to the one obtained by real-space identification ($r$-split) and thus has a better constraining power than DS performed in redshift space ($z$-split). Building a model for \textit{recon}-split is expected to be more challenging than for the other two identification scenarios. A tempting shortcut would be to build a model for $r$-split multipoles and compare it with \textit{recon}-split multipoles measured from real data. Although this approach seems to work on large scales, it could lead to significant biases in the inferred cosmological parameters below $\sim 20\,h^{-1}{\rm Mpc}$.

\begin{table*}
    \centering
    \caption{Comparison to Fisher forecasts for different summary statistics also based on the halo field.}
    \begin{tabular}{l c c c c c c c c c c r}
        \hline
        \hline
        Statistic & Scales & Redshifts & ${\rm \Omega_m}$ & $M_\nu$ & ${\rm \Omega_b}$ & $h$ & $n_s$ & $\sigma_8$  & Reference \\
        \hline
        ${\rm DS}^{\rm qq + qh}_{1+2+4+5}$ ($z$-split) & $10 < r < 150$ & $0.0$ & $\pm 0.01128$ & $\pm 0.05330$ & $\pm 0.0067$ & $\pm 0.06560$ & $\pm 0.04231$ & $\pm 0.02231$ & This work  \\
        ${\rm DS}^{\rm qq + qh}_{1+2+4+5}$ ($r$-split) & $10 < r < 150$ & $0.0$ & $\pm 0.00346$ & $\pm 0.05115$ & $\pm 0.00632$ & $\pm 0.05486$ & $\pm 0.03378$ & $\pm 0.01288$ & This work  \\
        Halo 2PCF & $10 < r < 150$ & $0.0$ & $\pm 0.04221$ & $\pm 0.39053$ & $\pm 0.01686$ & $\pm 0.20976$ & $\pm 0.24549$ & $\pm 0.11742 $ & This work  \\
        $B_0(k)$ & $k < 0.5$ & $0.0$ & $\pm 0.011$ & $\pm 0.054$ & $\pm 0.004$ & $\pm 0.039$ & $\pm 0.034 $ & $ \pm 0.014 $ & \cite{Hahn2020}  \\
        \textit{k}NN & $10 < r < 40$ & $[0.0, 0.5]$ & $\pm 0.0111$ &  $\pm 0.0925$ & $\pm 0.0029$ & $\pm 0.0273$ & $\pm 0.0206$ & $\pm 0.0108$ & \cite{Banerjee2021} \\
        MST(d,l,b,s) & $k < 0.5$ & $0.0$ & $\pm 0.036$ & $\pm 0.23$ & $\pm 0.0083$  & $\pm 0.073$ & $\pm 0.065$ & $\pm 0.067$ & \cite{Naidoo2022} \\
        Void 2PCF & $15 < r < 200$ & $0.0$ & $\pm 0.037$ & $\pm 0.13$ & $-$  & $\pm 0.089$ & $\pm 0.086$ & $\pm 0.067$ & \cite{Kreisch2021} \\
        Void-halo CCF & $15 < r < 200$ & $0.0$ & $\pm 0.027$ & $\pm 0.10$ & $-$  & $\pm 0.067$ & $\pm 0.066$ & $\pm 0.063$ & \cite{Kreisch2021} \\
        \hline
        \hline
    \end{tabular}
    \label{tab:constraints}
\end{table*}

\section{Discussion and conclusions} \label{sec:discussion_conclusions}

In this work, we have studied the cosmological constraining power of density-split clustering \citep[DS,][]{Paillas2021} in the context of the $\nu \Lambda$CDM model. This method consists in characterising the clustering of biased tracers as a function of environmental density, exploiting the sensitivity of each environment (density quintiles) to the cosmological parameters. DS offers an alternative to extract non-Gaussian information from a galaxy survey. The density field at small scales is highly non-Gaussian due to non-linear gravitational evolution, and therefore the power spectrum or the two-point correlation function (2PCF), which are measures of the variance of the density field, become incomplete descriptions of the galaxy distribution. DS is able to capture the missing information through a collection of correlation functions that are conditioned on environmental density, which naturally captures the non-Gaussian nature of the PDF.

We quantify the information content of DS through the Fisher matrix, estimated numerically from the halo catalogues of the Quijote suite of simulations \citep{Villaescusa2020}. We have found that DS improves the constraints on each cosmological parameter between a factor of $3$ and $8$, compared to the standard halo two-point correlation function.

In \citet{Paillas2021}, it was already shown that the cross-correlations between galaxies and DS quintiles could improve the constraints on the growth rate of structure by $30$ per cent over the 2PCF function analysis if the Gaussian streaming model \citep{Peebles1980, Fisher1995} was used to model the real-to-redshift space mapping. However, the analytical model presented in \citet{Paillas2021} relied on measurements of cross-correlation functions of real space galaxy catalogues from $\Lambda$CDM simulations, and their cosmological dependence was ignored in the analysis. This limits the amount of cosmological information that can be extracted to that of the real-to-redshift-space mapping. Here, we have shown for the first time that if we can model the full cosmological dependence of DS using N-body simulations, we can obtain much tighter constraints.

Moreover, we have presented the autocorrelations of the DS quintiles for the first time and have shown that they are also a valuable source of cosmological information, in addition to the DS cross-correlation functions. In particular, the quintile autocorrelations can recover some of the cosmological information that is lost when performing the density split in redshift space.

The Quijote simulations have allowed us to explore the sensitivity of DS clustering to different cosmological parameters, such as the sum of neutrino masses $M_{\nu}$. The combination of all DS quintiles places a constraint of $\sigma_{M_{\nu}} =  0.05330$ for a $(1\, h^{-1}{\rm Gpc})^3$ volume, assuming that we can model the redshift-space DS multipoles down to a scale of $10\,h^{-1}{\rm Mpc}$. Similarly, we obtain $\sigma_{\Omega_{\rm m}} = 0.01128$, $\sigma_{\Omega_{\rm b}}=0.0067$, $\sigma_h = 0.06560$, $\sigma_{\sigma_8}=0.02231$, and $\sigma_{n_s}=0.04231$, which corresponds to a factor of 3.7, 2.5, 3.2, 5.3, and 5.8 of improvement over the 2PCF, respectively. We note that our constraints are conservative, since the number density of resolved dark matter halos in the Quijote simulations is much lower than that expected in future galaxy surveys.

Our results are in line with forecasts from other summary statistics that aim at extracting non-Gaussian information from density fields. A natural approach is to include higher-order correlation functions or polyspectra. \citet{Hahn2020} found that the redshift-space halo bispectrum provides tighter constraints on the cosmological parameters of $\nu\Lambda$CDM, compared to the halo power spectrum. In particular, the bispectrum is five times better at constraining the sum of neutrino masses $M_{\nu}$, assuming that the bispectrum can be modelled up to $k_{\rm max} = 0.5\, h\,{\rm Mpc^{-1}}$. Including even higher-order correlations might tighten the cosmological constraints; however, even the full hierarchy of polyspectra may fail to contain all statistical information; see \cite{Carron_2011} for an example using log-normal fields. Moreover, the signal-to-noise ratio of higher-order moments decreases with the order of the correlators, and the computational complexity of higher-order statistics rises with the order of function chosen. Therefore, it is important to develop alternative statistics to the hierarchy of moments. 

Most alternative summary statistics exploit the environmental dependence of clustering, but differ on the particular definition of environment. \cite{Massara2022} showed that the marked power spectrum of the galaxy field can improve the constraints over the standard power spectrum by a factor of 3-6 for the $\nu \Lambda$CDM parameters. In their method, galaxies are weighted or `marked' with a function that depends on local density. Marks can be chosen so that low-density regions are up-weighted, which increases the sensitivity of the clustering to certain regions of the parameter space. As opposed to DS, where the density field is sampled around random centres, marked correlations use the positions of tracers to determine environment densities, and therefore their sensitivity to regions where there are no galaxies (such as void centres) may be different.

\citet{Uhlemann2020} showed that the one-point probability distribution function of counts-in-cells statistics provides particularly powerful constraints for $\Omega_{\rm m}$, $\sigma_8$ and $M_{\nu}$. They highlight the importance of combining information from different redshift bins in order to maximise information gain, which is something we have not explored in this work but could potentially be promising for DS. Moreover, given the low number density of our halo catalogues, we have not explored the additional information that the PDF might bring to DS statistics in full detail. We plan to study how complementary these two statistics are in future work.

\citet{Banerjee2021} used the k-nearest-neighbour (\textit{k}NN) distributions of haloes as a way to constrain cosmology. Validating their method with the Quijote halo catalogues, they found that the \textit{k}NN cumulative distribution functions improve the constraints on the cosmological parameters by roughly a factor of 4, using the scale range $10 < s < 40\,h^{-1}$Mpc and two redshift slices $z=0,0.5$. \citet{Naidoo2022} has analysed the information content of the minimum spanning tree (MST), the minimum weighted graph that connects a set of points without forming loops, finding that the MST breaks common parameter degeneracies in the $\nu \Lambda$CDM model, tightening the constraints on $M_{\nu}, h$, and $n_s$. 

One could also detect the positions in the cosmic web of tracers of different environments and use their statistics to constrain cosmology. For example, \cite{Kreisch2021} looked at the constraining power of cosmic void statistics, finding that the void size function, the void autocorrelation, and the void-halo cross-correlation functions provide tight constraints on $M_{\nu}$ on their own. Moreover, \citet{Bonnaire2022} used the eigenvalues of the tidal tensor to segment the cosmic web into nodes, filaments, walls, and voids, and used them to compute their respective power spectra in real space. In this paper, we have shown that cross-correlations between the halo field and the different environments add additional cosmological information to that of the autocorrelations (see Fig.~\ref{fig:likelihood_bridge}). Although the environment here is defined differently from \citet{Bonnaire2022}, we expect that similar gains could be achieved through the introduction of cross-correlation using their environment definition. Moreover, \citet{Bonnaire2022} assumed that the real space positions of the tracers were known when identifying environments, but did not analyse the impact that identifying environments in redshift space could have on the resulting cosmological constraints.

Table~\ref{tab:constraints} summarises the constraining power of different summary statistics found using the dark matter halos of the Quijote suite of simulations. We do not include studies based on the dark matter or galaxy field, since a one-to-one comparison would not be possible. It shows how DS can obtain state-of-the-art constraints on the cosmological parameters $\Omega_{\rm m}$, $M_\nu$, and $n_s$ while still obtaining competitive constraints on the remaining parameters. Rather than advocating for a particular summary statistic, we highlight the possibility of complementing these different probes, exploiting the degeneracy-breaking power that each of them has to offer. We caution the reader that our reported cosmological constraints, especially those for $\Omega_{\rm b}$, $h$, and $n_s$, should not be taken at face value as precise parameter forecasts for galaxy surveys, since they rely on the estimation of numerical derivatives that could be considered as not being fully converged (see Fig.~\ref{fig:convergence_fisher}. Instead, they should be interpreted to assess the relative improvement in constraining power between different summary statistics that operate on the same data set.

We have shown that the DS clustering statistics depend on whether the density environments are defined in real or redshift space. Real-space identified quintiles yield better constraints for all cosmological parameters, in particular $\Omega_{\rm m}$ and $\sigma_8$, and indeed in \cite{Paillas2021} it was shown that if one has access to the real-space galaxy positions to identify the quintiles in this way, it is possible to model the real-to-redshift space mapping of the DS cross-correlation functions analytically using the Gaussian streaming model down to $\sim 15\,h^{-1}$Mpc. However, galaxy catalogues in real space are not immediately available in observations, and one would have to rely on reconstruction algorithms to approximately remove RSD from galaxies \citep{Nadathur2019c}. But, as shown in Sect.~\ref{sec:bias_reconstruction}, reconstruction algorithms could potentially introduce systematic errors in the inferred cosmological parameters when not modelled appropriately, which would then need to be added to the total error budget.

When presenting the main cosmological constraints of our analysis, we have put aside the complications related to theoretical modelling and implicitly assumed that we have access to a model that can perfectly match the measurements down to $10\,h^{-1}{\rm Mpc}$. An analytical prediction of how the multipoles of DS statistics change with cosmology is a challenging task. We plan to work on a simulation-based model to allow for a comparison between simulations and data, which will be presented in a future paper. This framework could potentially allow us to directly emulate the redshift-space DS multipoles, without the need for reconstruction. Moreover, we have focused here on DS statistics for dark matter halos, but we will work on simulation-based models for the DS statistics of galaxies. We expect DS to set tight constraints on environment-based assembly bias  \citep{Xu_2021}.

We note that since the different samples obtained through DS are expected to share the same sample variance, they can also make use of sample variance cancellation techniques such as proposed in~\cite{McDonald2008:0810.0323v1} and~\cite{Seljak2008:0807.1770v1}. In fact, part of the gain in signal-to-noise we obtained over the standard 2PCF analysis might be related to this effect. However, sample variance cancellation can only meaningfully contribute to the signal-to-noise if the shot noise contribution is small, which is not the case for the Quijote simulations. However, DS could be a promising analysis technique to exploit sample variance cancellation in a future high-density sample such as DESI-BGS \citep{Zarrouk2022}. 

Zero-biased tracers have been shown to be a promising way to achieve optimal constraints on primordial non-Gaussianity~\citep{Castorina2018:1803.11539v1}. Since it is basically impossible to obtain zero-biased tracers through colour or magnitude cuts, DS again might provide a useful tool for such studies. 

Relativistic effects can only be analysed in the cross-correlation of differently biased tracers, the signal itself being proportional to the difference in galaxy bias~\citep{Yoo2010:1009.3021v1,Bonvin2011:1105.5280v3,Challinor2011:1105.5292v2}. DS might prove useful for such studies, given the wide range in galaxy bias accessible with this technique.

Ongoing and upcoming large-area surveys, such as DESI \citep{desi}, Euclid \citep{euclid}, and Roman Space Telescope \citep{roman}, will offer unprecedented statistical precision for galaxy clustering, due to their large volume coverage and galaxy number density. A tremendous amount of information from these Stage-IV experiments will be available in the mildly non-linear regime, where the density field is non-Gaussian. Methods that can grant access to higher-order statistical information beyond two-point statistics, such as DS, will thus play a key role in extracting cosmological information that cannot be readily accessed with the power spectrum. This will require percent-level precision from the modelling side, while ensuring that the models can circumvent the observational systematic effects that will be inherent to these datasets. A noteworthy difficulty compared to the idealised scenario of this paper is that one will need to account for the selection function of the survey when estimating the overdensities around the random centres.

\section*{Acknowledgements}
We would like to thank Elena Massara, Zhongxu Zhai, Baojiu Li, Ravi Sheth, Ariel S\'anchez, and Oliver Philcox for helpful discussions. We acknowledge the use of \textsc{matplotlib} \citep{matplotlib}, \textsc{scipy} \citep{scipy}, and \textsc{astropy} \citep{astropy2013, astropy2018} throughout the course of this work. This research was enabled in part by support provided by Compute Ontario (computeontario.ca) and the Digital Research Alliance of Canada (alliancecan.ca). Research at Perimeter Institute is supported in part by the Government of Canada through the Department of Innovation, Science and Economic Development Canada and by the Province of Ontario through the Ministry of Colleges and Universities. SN acknowledges support from an STFC Ernest Rutherford Fellowship, grant reference ST/T005009/2. YC acknowledges the support of the Royal Society through the award of a University Research Fellowship and an Enhancement Award. This project has received funding from the European Research Council (ERC) under the European Union’s Horizon 2020 research and innovation program (grant agreement 853291). FB is a University Research Fellow.  This work used the DiRAC@Durham facility managed by the Institute for Computa- tional Cosmology on behalf of the STFC DiRAC HPC Facility (www.dirac.ac.uk).

For the purpose of open access, the authors have applied a CC BY public copyright licence to any Author Accepted Manuscript version arising.

\section*{Data Availability Statement}

The source code and data needed to generate the figures in this manuscript are available at \url{https://github.com/epaillas/densitysplit-fisher}.



\bibliographystyle{mnras}
\bibliography{references} % if your bibtex file is called example.bib

%%%%%%%%%%%%%%%%%%%%%%%%%%%%%%%%%%%%%%%%%%%%%%%%%%

%%%%%%%%%%%%%%%%% APPENDICES %%%%%%%%%%%%%%%%%%%%%

\appendix

\section{The impact of mixing quintiles when estimating overdensities in redshift space} \label{ap:r_vs_z}
In this section, we examine the contribution to the quadrupole of quintile autocorrelations in terms of the signal coming from random centres that have been correctly identified in redshift space, and those that have been misidentified.

Let us begin by defining the set of correctly identified random points for $\mathrm{DS}_i$ as
\begin{equation}
    \rm Z \cap \rm R = \left \{ \mathbf{x} \in \left( \mathrm{DS}_i^\mathrm{Z} \cap \mathrm{DS}_i^\mathrm{R}  \right) \right \},
\end{equation}
where superscript Z and R, denote redshift- and real-space identification, respectively. We denote those incorrectly identified as
\begin{equation}
    \rm Z \notin \rm R = \left \{ \mathbf{x}: \mathbf{x} \in \mathrm{DS}_i^\mathrm{Z}, \mathbf{x} \notin  \mathrm{DS}_i^\mathrm{R}  \right \}.
\end{equation}
For a given density split quintile, $\mathrm{DS}_i$, we separate the contribution to the quadrupole from the two sets as
\begin{align} \nonumber
    \label{eq:quadrupole_correct_incorrect}
    \xi_2^\mathrm{qq} = &\left(\frac{|\rm Z \cap \rm R|}{N_{\rm random}}\right)^2\xi_2^{\rm Z \cap \rm R} +  \left(\frac{|\rm Z \notin \rm R|}{N_{\rm random}}\right)^2 \xi_2^{\rm Z \notin \rm R} \\ &+ 2 \frac{|\rm Z \cap \rm R| |\rm Z \notin \rm R|}{N_{\rm random}^2} \xi_2^{\rm Z \cap \rm R, \rm Z \notin \rm R }
\end{align}
where $|Z \cap R|$ and $|Z \notin R|$ are the number of points correctly and incorrectly identified, respectively. The first term in Eq.~(\ref{eq:quadrupole_correct_incorrect}) quantifies the anisotropy resulting from missing random centres that have not been correctly identified, the second term represents the contribution of anisotropies present in the random centres that have been incorrectly added, whereas the last term quantifies the cross-correlation between those centres that have been correctly identified and those that have been added.

\begin{figure}
    \centering
    \includegraphics[width=0.47\textwidth]{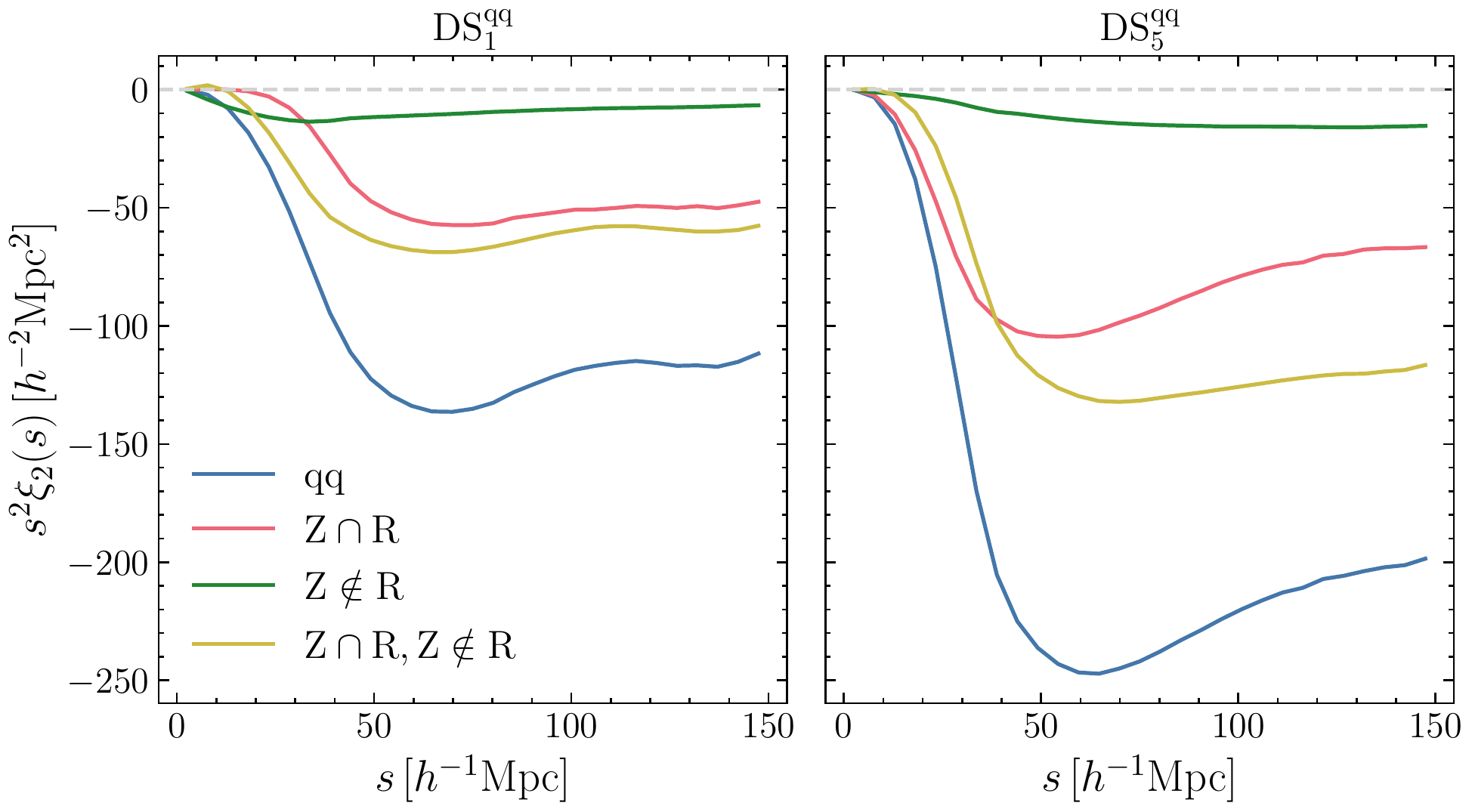} 
    \caption{The contribution from correctly ($\rm Z \cap \rm R$) and incorrectly classified ($\rm Z \notin \rm R$) random points to the quadrupole of autocorrelations. We show both the effect for $\mathrm{DS}_1$ (left) and $\mathrm{DS}_5$ (right), estimated for only one realization of the fiducial Quijote simulations. \href{https://github.com/epaillas/densitysplit-fisher/blob/master/FigureA1.py}{\faFileCodeO}}
    \label{fig:quadrupole_contributions}
\end{figure}

Figure \ref{fig:quadrupole_contributions} shows the contribution of each term in Eq.~(\ref{eq:quadrupole_correct_incorrect}). For both  $\mathrm{DS}_1$  and $\mathrm{DS}_5$, all terms contribute to the overall squashing of the autocorrelation. 

\section{Correction to the covariance matrix}
\label{ap:covariance_correction}

We now consider how to debias the errors on model parameters obtained from our Fisher-matrix-based procedure given that the covariance matrix $C$ is derived from simulations [Eq.~(\ref{eq:covariance})] and hence represents a random draw from a Wishart distribution. The following derivation forms part of the analysis presented in \cite{Percival2022} -- see Section 3.1 of that paper -- but was not considered in our context. Consider a Gaussian Fisher information matrix with a covariance matrix $C$,
\begin{equation}  
  F_C =
  {\frac{\partial {\bm d}}{\partial {\bm\theta}}}^\top C^{-1}
  {\frac{\partial {\bm d}}{\partial {\bm\theta}}}\,,
\end{equation}
where $C$ is calculated using $n_{\rm sim}$ simulations, effectively being drawn from a Wishart distribution 
\begin{equation}
  f(C|\Sigma)=f_W(C|\Sigma/(n_{\rm sim} - 1),n_{\rm sim} - 1)\,,
\end{equation}
with expected (true) value $\Sigma$. ${\bm\theta}$ are the model parameters and ${\bm d}$ the data vector.

The true Fisher matrix would be
\begin{equation}  \label{eq:Ftrue}
  F_\Sigma =
  {\frac{\partial {\bm d}}{\partial {\bm\theta}}}^\top \Sigma^{-1}
  {\frac{\partial {\bm d}}{\partial {\bm\theta}}}\,,
\end{equation}
and unbiased model parameter error estimates could be obtained by inverting this relation if we knew $\Sigma$. That is, for the variances quoted on the model parameters, we want $F_\Sigma^{-1}$, but we use an estimator with mean $\langle F_C^{-1}\rangle=\langle (MC^{-1}M^\top)^{-1}\rangle$, where $M\equiv\frac{\partial
  {\bm d}}{\partial {\bm\theta}}^\top$ is a $n_\theta\times n_{\rm bins}$ matrix, with $n_\theta$ the number of model parameters, and $n_{\rm bins}$ the size of the data vector.

A property of the Fisher matrix is that
\begin{align} \nonumber
  &f(\,(MC^{-1}M^\top)^{-1}|\Sigma) =\\
  &f_W\left((MC^{-1}M^\top)^{-1}\left|
    \frac{(M\Sigma^{-1}M^\top)^{-1}}{n_{\rm sim}-1},
        n_{\rm sim}-n_{\rm bins}+n_\theta-1\right.\right) 
\end{align}
(see Theorem 3.2.11 in \citealt{Muirhead1982}).

From the expectation of a Wishart-distributed variable [for
$f_W(C|R,\nu)$, $\langle C\rangle=\nu R$] we can write
down
\begin{equation}
  \langle (MC^{-1}M^\top)^{-1}\rangle = \frac{n_{\rm sim}-n_{\rm bins}+n_\theta-1}{n_{\rm sim}-1}F_\Sigma^{-1}\,.
\end{equation}
Thus, we see that we need to use $C'$ rather than $C$, where
\begin{equation}  \label{eq:fac}
  C' = \frac{n_{\rm sim}-1}{n_{\rm sim}-n_{\rm bins}+n_\theta-1}C\,,
\end{equation}
in order to obtain an unbiased estimator for $F_\Sigma^{-1}$.
We note that this is close to the \citep{Hartlap2007} factor, which would give
\begin{equation}  \label{eq:hartlap}
  C' = hC = \frac{n_{\rm sim}-1}{n_{\rm sim}-n_{\rm bins}-2}C\,,
\end{equation}
except where $n_\theta$ is large. This can be easily understood: the factor in Eq.~(\ref{eq:fac}) corrects for skewness in both the inversion of $C$ to give the Fisher matrix and the subsequent inversion of the Fisher matrix to obtain parameter constraints. The Hartlap factor only corrects for the first of these inversions. Thus, the correct factor looks like the Hartlap factor, when the number of model parameters is small, and no additional skewness is introduced by the inversion of the Fisher matrix. However, if $n_\theta\sim n_{\rm bins}$, the skewness of the second inversion cancels that of the first, and the factor reduces to unity. 

\section{Assessing the Gaussianity of the density-split likelihood} 
\label{ap:gaussianity_likelihood}

\begin{figure*}
    \centering
     \includegraphics[width=1.\textwidth]{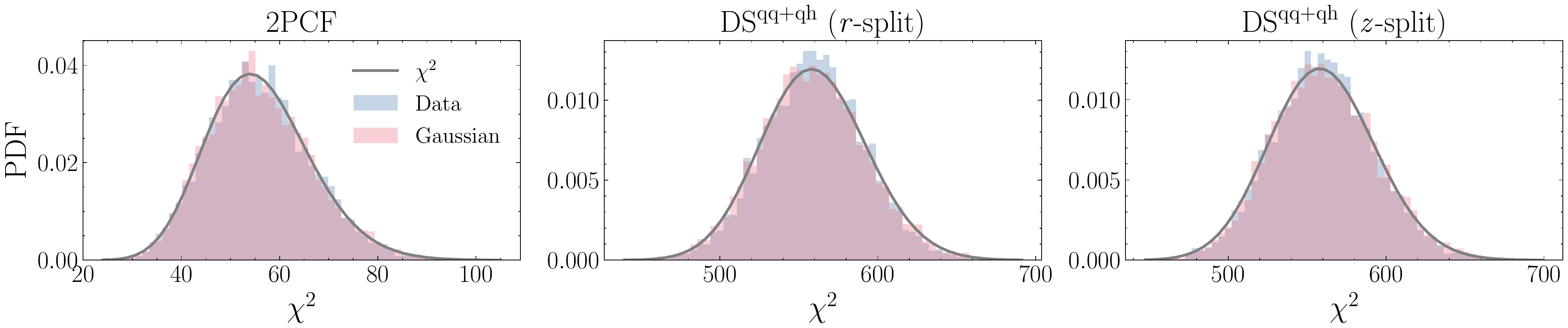} 
    \caption{A qualitative assessment of the Gaussianity of the likelihoods for the 2PCF (left), DS identified in real space (middle) and DS identified in redshift space (right). The colored histograms show the distribution of $\chi^2$ values, as measured from the Quijote simulations (blue) and a multivariate Gaussian distribution with the same mean and covariance as the simulations (pink). The solid line shows a theoretical $\chi^2$ distribution with degrees of freedom set to the number of pair separation bins. \href{https://github.com/epaillas/densitysplit-fisher/blob/master/FigureC1.py}{\faFileCodeO}}
    \label{fig:gaussian_test}
\end{figure*}

In this section, we check that the likelihood of DS statistics is distributed as multivariate Gaussian, following the analysis in \cite{Friedrich2021}. We first compute the $\chi^2$ value of the summary statistic measured in each of the fiducial simulations
\begin{equation}
    \chi^2_i = \left(\bm{d_i}(\mathbf{s}) - \bm{\bar{d}}(\mathbf{s})\right)^\top C^{-1}  \left(\bm{d_i}(\mathbf{s}) - \bm{\bar{d}}(\mathbf{s})\right),
\end{equation}
where $\bm{d_i}$ represents the value of the summary statistic for the $i$-th fiducial simulation evaluated at the pair separation vector $\mathbf{s}$, $\bm{\bar{d}}(\mathbf{s})$ is the average of the summary statistic over all fiducial simulations at the pair separation vector $\mathbf{s}$, and $C$ is the covariance matrix estimated from all the fiducial simulations.

If the likelihood of the summary statistic is Gaussian distributed, the $\chi^2_i$ values should also follow a $\chi^2$ distribution with degrees of freedom determined by the number of pair-separation bins.

Furthermore, if the likelihood is Gaussian, the distribution of $\chi^2_i$ should also be very close to that of sampling from a multivariate Gaussian with a mean given by $\bar{d}$ and the covariance measured from the simulations.

In Fig.~\ref{fig:gaussian_test}, we show how the 2PCF and DS statistics $\chi^2_i$ calculated from the data follow a very similar $\chi^2$ distribution as that of the random samples generated from a multivariate Gaussian.

\section{Convergence of Fisher forecasts} 
\label{ap:convergence_tests}

\begin{figure*}
    \centering
    \begin{tabular}{cc}
      \includegraphics[width=0.4\textwidth]{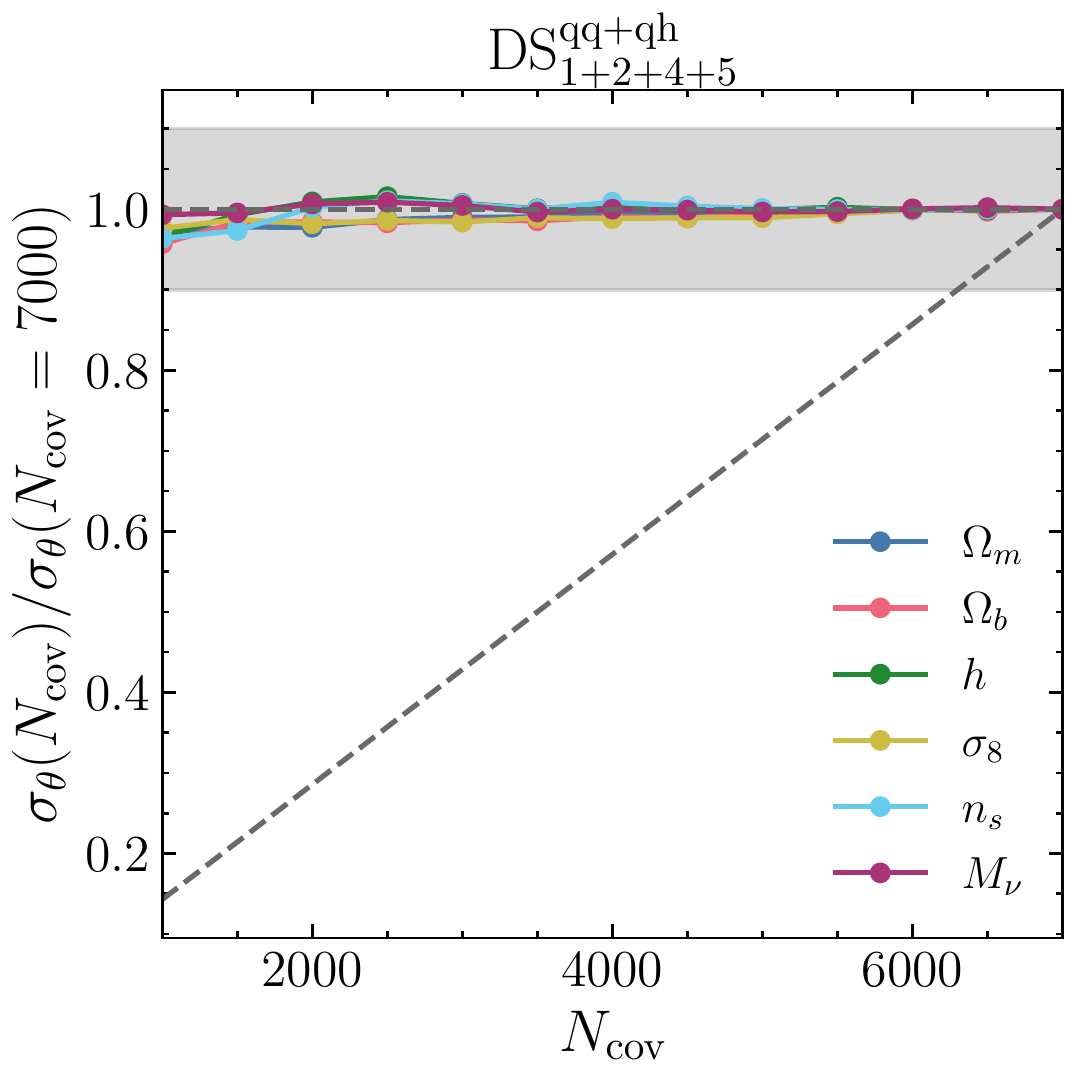}  & \includegraphics[width=0.4\textwidth]{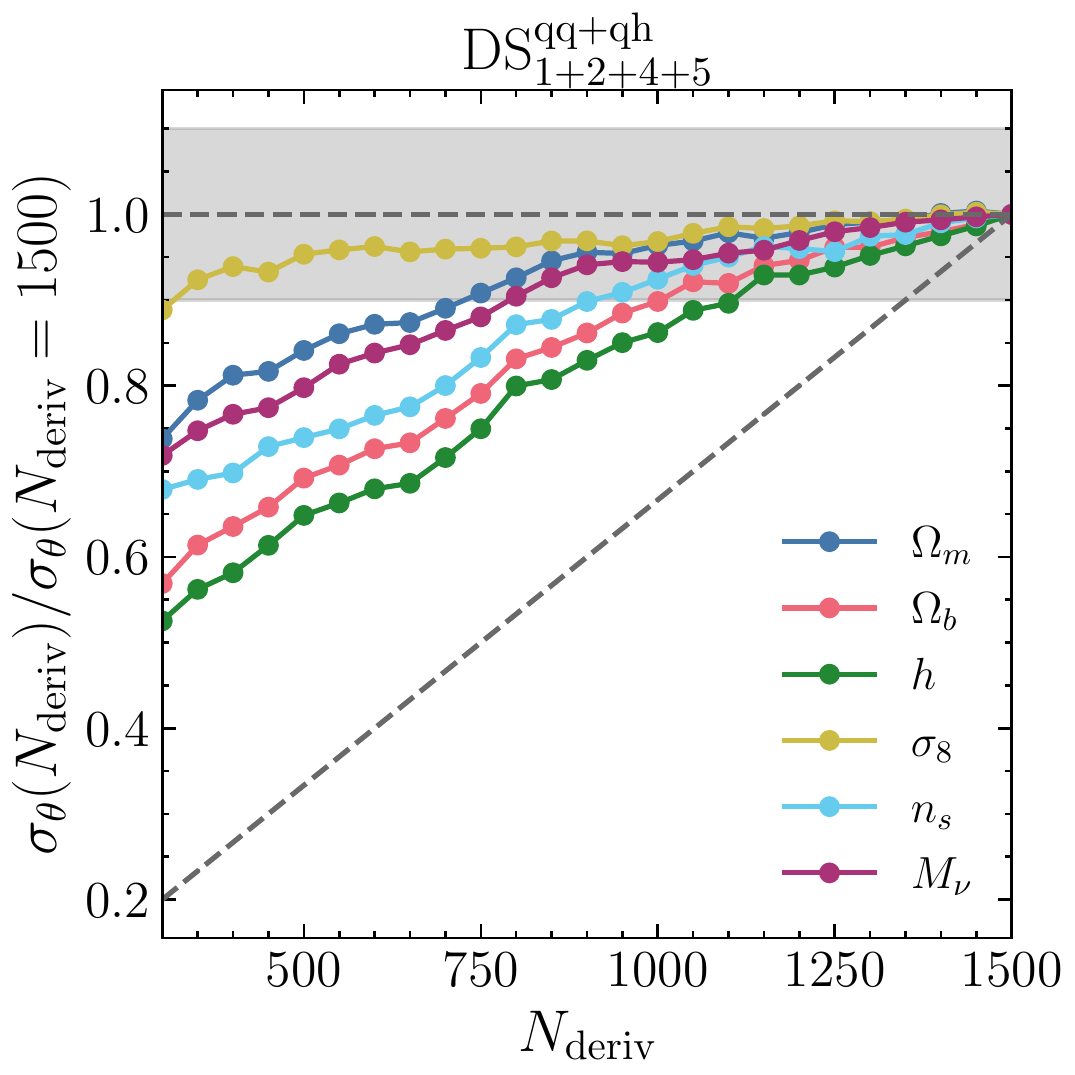}
    \end{tabular}
    \caption{Convergence of the constraints estimated through the Fisher matrix. The left-hand panel shows the 1-$\sigma$ errors on each model parameter as a function of the number of mocks used to estimate the covariance matrix, where the errors are normalized by the default case when $N_\mathrm{cov} = 7000$. The right-hand panel shows the same convergence tests but for the number of mocks used to estimate the derivatives, where the errors are normalized by the default case $N_\mathrm{deriv} = 1500$. The grey shaded bands show regions where the agreement is better than 10 per cent. The dashed lines show the variation in number of mocks compared to the default cases, either $N_{\rm cov}/7000$ (left) or $N_{\rm deriv}/1500$ (right). \href{https://github.com/epaillas/densitysplit-fisher/blob/master/FigureD1a.py}{\faFileCodeO} \href{https://github.com/epaillas/densitysplit-fisher/blob/master/FigureD1b.py}{\faFileCodeO}}
    \label{fig:convergence_fisher}
\end{figure*}

The results presented in Sect.~\ref{subsec:ds_vs_tpcf} are based on Fisher matrices estimated using a finite number of mocks. There are two ingredients that are needed to calculate them: the derivatives of the multipoles with respect to the cosmological parameters, and the covariance matrix of the multipoles in the fiducial cosmology.

Figure \ref{fig:convergence_fisher} shows how do the inferred errors on the cosmological parameters change as we increase the number of simulations used to estimate the derivatives and the covariance matrix. The results are expressed in terms of the 1-$\sigma$ errors of the model parameters, compared to the limiting case of 1500 simulations for the derivatives and 7000 simulations for the covariance matrix, as adopted in the paper.

For the covariance matrix, a tight convergence is guaranteed even when using a relatively small number of simulations: the errors on the parameters change by less than 2 per cent when using between 2000 and 4000 realisations, and by less than 1 per cent when using more than 4000 realisations.

The number of mocks that is used to estimate the derivatives has a strong impact on the inferred errors on the parameters. Only when using more than $\sim 1100$ realisations, we can expect fluctuations in the errors of less than 10 per cent for all parameters.

We have also explicitly checked that a similar convergence is achieved for the 2PCF multipoles, as well as for the density-split multipoles identified in real space, both for Quijote and the Gaussian mocks presented by the end of Sect.~\ref{subsec:ds_vs_tpcf}.

%%%%%%%%%%%%%%%%%%%%%%%%%%%%%%%%%%%%%%%%%%%%%%%%%%

% Don't change these lines
\bsp	% typesetting comment
\label{lastpage}
\end{document}